\documentclass{article}

\usepackage{arxiv}

\usepackage{graphicx}
\usepackage[utf8]{inputenc} 
\usepackage{indentfirst,csquotes}

\usepackage[T1]{fontenc}    
\usepackage{hyperref}       
\usepackage{url}            
\usepackage{booktabs}       
\usepackage{amsfonts}       
\usepackage{nicefrac}       
\usepackage{microtype}      
\usepackage{lipsum}
\usepackage{graphicx}
\usepackage{amsmath}
\graphicspath{ {./images/} }

\title{Fundamental Limitations of Terahertz Quarter-Wave Plates Based on High-Contrast Dielectric Gratings}

\author{
 Oleg Kameshkov \\
  ARC Centre of Excellence for Transformative Meta-Optical Systems (TMOS),  \\
  Research School of Physics, The Australian National University\\
  Canberra, ACT 2601, Australia \\
  \texttt{Oleg.Kameshkov@anu.edu.au} \\
   \And
 Rajour Tanyi Ako \\
 ARC Centre of Excellence for Transformative Meta-Optical Systems (TMOS) \\
  and Functional Materials and Microsystems Research Group, RMIT University,\\
  Melbourne, Australia \\
   \And
Madhu Bhaskaran \\
 ARC Centre of Excellence for Transformative Meta-Optical Systems (TMOS) \\
  and Functional Materials and Microsystems Research Group, RMIT University,\\
  Melbourne, Australia \\
   \And
Ilya Shadrivov \\
  ARC Centre of Excellence for Transformative Meta-Optical Systems (TMOS),  \\
  Research School of Physics, The Australian National University\\
  Canberra, ACT 2601, Australia \\
}

\begin{document}

 \maketitle

\begin{abstract}
We present a systematic study of high-contrast dielectric gratings operating as broadband quarter-wave plates in the terahertz range. Using higher-order effective medium theory, we identify an achromatic operating regime characterized by a parabolic-like frequency dependence of the phase retardation. This framework reveals a fundamental bandwidth limit imposed by single-mode operation in one-dimensional rectangular gratings. We show that Fabry–Perot resonances and the higher-order modes set intrinsic constraints on achievable performance.
The theoretical predictions are quantitatively validated by rigorous full-wave simulations and terahertz time-domain spectroscopy experiments. Our results reveal clear physical design rules and intrinsic performance limits for broadband terahertz dielectric grating waveplates. 
\end{abstract}

\section{Introduction}
\label{sec:Introduction}
\noindent
Terahertz (0.1 - 10 THz) time-domain spectroscopy (THz-TDS) is a valuable technique  for non-destructive material analysis and for retrieving optical parameters such as the refractive index and absorption coefficient. However, detectors of conventional THz-TDS are typically sensitive to a single polarization and are therefore limited to measuring amplitude and phase information for a given polarization~\cite{withayachumnankul_fundamentals_2014, neu_tutorial_2018, koch_terahertz_2023}. As a result, polarization effects cannot be properly resolved without specialized experimental configurations, and THz-TDS is most commonly applied to the investigation of isotropic materials \cite{pfleger_advanced_2014, choi_terahertz_2022}.

Over time, many research groups have developed their polarimetric THz-TDS systems that would inherit the broadband capabilities of conventional setups  while enabling full-polarization state analysis \cite{neshat_improved_2012, dong_polarization_2009, dong_measurement_2010, gong_cross-polarization_2011, chen_introduction_2022, park_broadband_2024}. These THz time-domain polarimeters are typically realized by introducing either polarizers or a combination of polarization converters and waveplates into a standard THz-TDS system. While the former approach has been successfully implemented and enables analysis of anisotropic media and circular dichroism  \cite{choi_chiral_2022, kan_enantiomeric_2015, pfleger_advanced_2014}, it suffers from low performance and limited feasibility for integration into frequency-domain techniques.

The latter approach is hindered by the lack of naturally occurring materials that exhibit both low absorption and high birefringence in the terahertz range, which are required for conventional polarization optics such as waveplates. For example, the birefringence of quartz at 1~THz is approximately 0.046, while its absorption coefficient is \~ 0.2~cm$^{-1}$. Several groups have developed achromatic multilayer waveplates by bonding quartz plates together. However, their total thickness typically reaches several millimeters, resulting in high losses, high cost, and poor integrability \cite{masson_terahertz_2006, chen_terahertz_2013, zhang_terahertz_2021, zhang_general_2023}. Consequently, this classical approach does not meet the demands of broadband THz systems.  

Alternatively, electro-optic techniques have been explored to realize time-domain polarimetry without the need for external optical elements \cite{xu2020terahertz}. However, these methods suffer from mechanical rotation constraints, complex calibration procedures, and a limited angular sensitivity of only a few degrees.

Broadband polarization control of terahertz radiation remains a significant challenge. The most common polarization control tools are quarter-wave and half-wave plates. They convert linear polarization into circular and orthogonal polarization states, respectively, and they are typically fabricated from anisotropic materials and operate based on birefringence effect. Various strategies have been proposed to achieve achromatic THz waveplates  \cite{qiu_terahertz_2018, petrov_design_2022, gong_research_2023, zhang_review_2024}, including designs based on subwavelength gratings \cite{zhang_achromatic_2015-1}, metamaterials  \cite{cong_highly_2014}, prism-type wave plate \cite{kawada_achromatic_2014}, natural birefringent materials \cite{masson_terahertz_2006}, liquid crystals \cite{wang_broadband_2015}, electro-optic polymers \cite{zhang_tunable_2020}, or their combinations \cite{chen_highly_2019}. Each method has its advantages and disadvantages. For example, metal-based structures have excellent performance at low THz frequencies, but multilayer configurations designed to enhance bandwidth typically suffer from increased reflection and internal losses. Liquid crystal based solutions face low birefringence, which results in bulky and absorptive designs, and they are also sensitive to temperature. On the other hand, dielectric based waveplates seem to be more attractive due to their superior transmission and low insertion loss, but still lack sufficient bandwidth for broadband applications. Furthermore, achieving a large phase difference between orthogonal polarizations remains difficult when using low-refractive-index materials.

One of the simplest ways to introduce artificial birefringence is to fabricate a one-dimensional (1D) {\em subwavelength} dielectric grating on a surface, in which waves polarized parallel and perpendicular to the grooves experience different phase retardations. This well-established approach has enabled the realization of broadband achromatic waveplates across the visible \cite{yi_broadband_2003}  and infrared \cite{liu_highly_2019, mutlu_experimental_2012, delacroix_design_2012, nordin_broadband_1999} spectral ranges.  In the terahertz regime, 1D dielectric gratings have also been employed for the development of phase shifters and waveplates. One of the research direction focuses on waveplates operating in the low-THz range, often using low-refractive-index 1D gratings fabricated by 3D printing or milling techniques \cite{dong_wideband_2019, rohrbach_3d-printed_2021, yuan_terahertz_2021, guan_terahertz_2022, yuan_high-transmission_2025, jackel_achromatic_2022}. In these studies, researchers match fundamental and higher-order modes in sub-diffraction gratings to suppress phase dispersion and achieve broadband polarization control \cite{guan_terahertz_2022} or dual-band polarization control \cite{yuan_terahertz_2021}. Another research direction consists of numerical studies investigating optimal grating designs  \cite{sun_achromatic_2012, zhang_design_2025, zhang_analysis_2024} as well as fabrication techniques \cite{saha_imprinted_2010, zhang_thin-form_2013}. For example, in 2025, Ji Zhang et.el \cite{zhang_design_2025} proposed a broadband tri-layer grating design, while Saha Shimul et.al. \cite{saha_imprinted_2010} demonstrated  imprinted high aspect ratio artificial dielectric quarter wave plates on polymers for operation at discrete THz frequencies.

Further research has focused on expanding the operational frequency range and improving grating performance through  various strategies \cite{zhang_achromatic_2015-1, chen_artificial_2016, mu_broadband_2019, wu_ultrabroadband_2022, ayyagari_broadband_2024, zhu_extend_2024}. In 2015, Banghong Zhang et al. \cite{zhang_achromatic_2015-1} introduced a design method based on second-order effective medium theory to create a broadband quarter-wave plate using a single-layer rectangular dielectric grating. In 2016, Meng Chen \cite{ chen_artificial_2016} demonstrated a chirped gradient grating and applied it as a broadband half-wave plate with improved birefringence and reduced dispersion. More recently, in 2024, Surya Revanth Ayyagari et al. \cite{ayyagari_broadband_2024} developed broadband half- and quarter-wave plates based on dielectric gratings with trapezoidal groove profiles and an overall sinusoidal grating shape, providing anti-reflection characteristics and suppression of Fabry-Perot effects.

A final area of active research involves combining dielectric gratings with other materials or metamaterials to achieve specialized functionalities  \cite{jiang_ultrawide_2023, xu_dispersioncompensated_2024}.  
In 2023, Songlin Jiang \cite{jiang_ultrawide_2023} combined a dielectric subwavelength rectangular grating with liquid crystals to realize a tunable phase shifter operating as quarter- and half-wave plates at discrete frequencies.  In 2024, Shi-Tong Xu \cite{xu_dispersioncompensated_2024} implemented dispersion compensation by pairing dielectric and metallic gratings exhibiting positive and negative phase-dispersion slopes, respectively. This approach flattened the overall dispersion and enabled broadband phase shifts.

Despite numerous studies on one-dimensional dielectric gratings in the terahertz range, the question of the fundamental performance limits of these structures remains open. In this paper, we systematically study the potential of one-dimensional rectangular dielectric gratings as broadband quarter-wave plates operating in the single-mode regime in the terahertz range. The paper is organized as follows. First, we introduce the higher-order effective medium theory to describe the grating and its interaction with an incident wave. We use this effective medium theory to optimize the structure parameters to identify maximum bandwidth of the quarter-wave plate at a frequency of interest. We then analyze the limitations of the theoretical model and compare it with rigorous numerical simulations. We then fabricate the structure and perform experimental demonstration of the silicon grating quarter-wave plate. Finally, we summarize our findings and discuss future perspectives of using one-dimensional dielectric gratings as wave plates in the terahertz range.
\section{Theoretical framework}

\subsection{Effective medium theory}
\begin{figure}[htp]
    \centering
    \includegraphics[width=6cm]{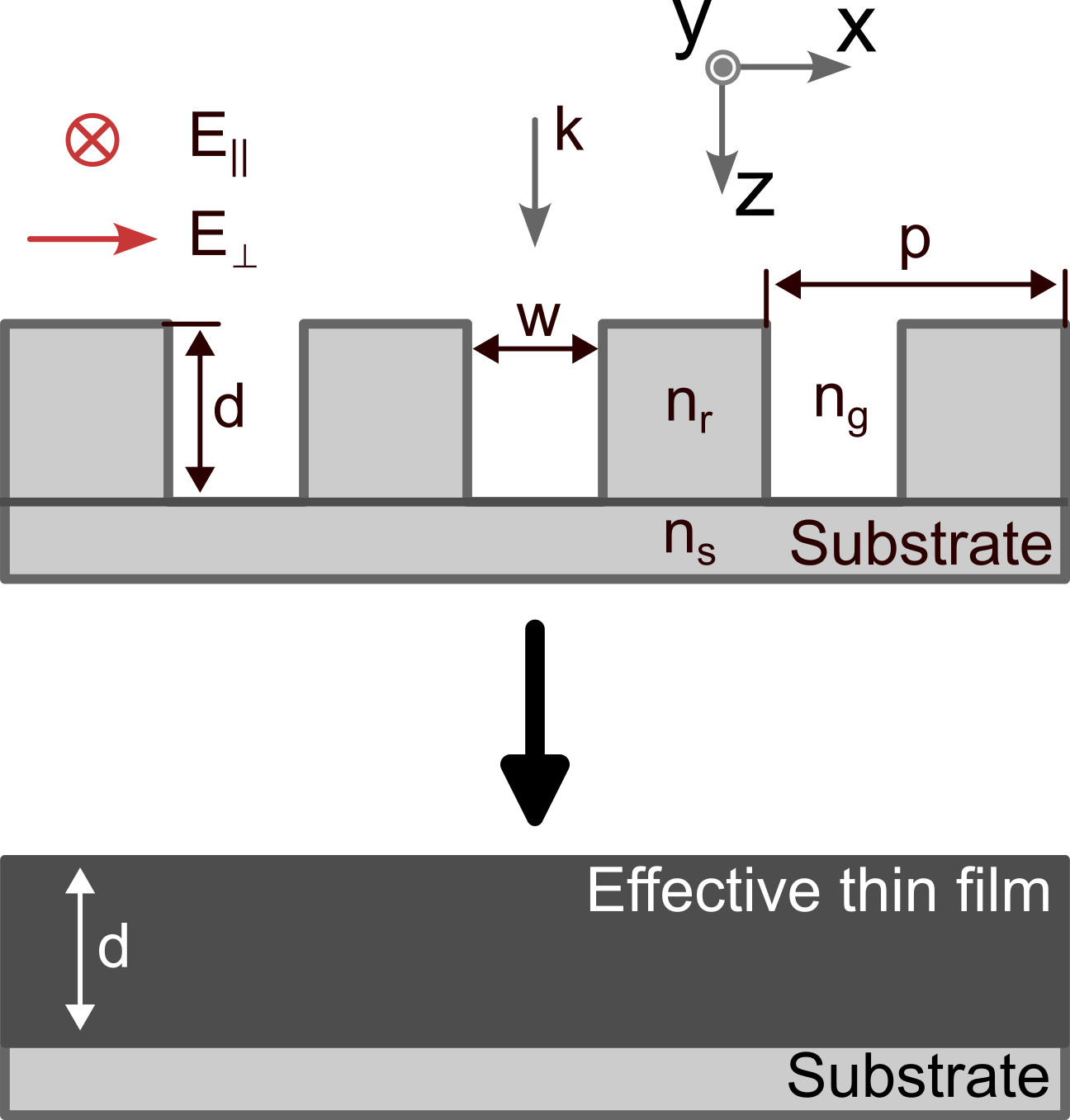}
    \caption{Rectangular dielectric grating on a substrate and its representation as a layer of an effective medium on a substrate.}
    \label{fig:Figure_1}
\end{figure}

We first review the effective medium formalism applicable to the description of a dielectric grating with rectangular grooves shown in Fig.~\ref{fig:Figure_1}.   The grating consists of equally spaced parallel grooves and ridges with refractive indices $n_g$ and $n_r$, respectively. It is assumed to be infinite in the $y$-direction and periodic in the $x$-direction. The grating period, groove width, ridge width, and depth are denoted by $p$, $w$, $r = p - w$, and $d$, respectively. The refractive index of the substrate is denoted by $n_s$, while the surrounding medium is assumed to be identical to the groove medium.
In this method, the grating layer is replaced by a homogeneous layer with an effective refractive index $n_{eff}$,  which lies between $n_g$ and $n_r$, as schematically shown in Fig.~\ref{fig:Figure_1}. This simplification reduces the problem of grating analysis to that of a planar multilayer system. Rytov \cite{rytov_electromagnetic_1956}  examined this problem by considering a plane wave propagation through a structure composed of alternate dielectric layers and derived transcendental equations for the effective refractive indices corresponding to the polarizations parallel and perpendicular to the grating grooves.
For parallel polarization, the effective refractive index  $n_{||}$ satisfies: 
\begin{equation} \label{eq:TE_Eff_med}
\begin{split}
    \sqrt{n_g^2-n_{||}^2} \cdot tan \left[\frac{\pi p}{\lambda} (1-F)\sqrt{n_g^2-n_{||}^2} \right] =\\= - \sqrt{n_r^2-n_{||}^2} \cdot tan \left[\frac{\pi p}{\lambda}F\sqrt{n_r^2-n_{||}^2} \right]
\end{split}
\end{equation}
whereas for perpendicular polarization, the effective refractive index $n_{\perp}$ satisfies
\begin{equation} \label{eq:TM_Eff_med}
\begin{split}
\frac{\sqrt{n_g^2-n_{\perp}^2}}{n_g^2} \cdot tan \left[ \frac{\pi p}{\lambda} (1-F)\sqrt{n_g^2-n_{\perp}^2} \right ] 
\\
= - \frac{\sqrt{n_r^2-n_{\perp}^2}}{n_r^2} \cdot tan \left[ \frac{\pi p}{\lambda}F\sqrt{n_r^2-n_{\perp}^2} \right]
\end{split}
\end{equation}
Here, $F=r/p$ denotes  the fill factor. These transcendental equations have an infinite number of solutions due to the periodic nature of the tangent function. These solutions corresponds to different modes excited inside the ridges under normal incidence characterized by a propagation constant  $\beta_m$, where the effective refractive index of the $m$-th mode is given by  $n_{eff}^m=\beta_m\cdot\lambda / 2\pi$.  

However, in most practical cases, only the lowest-order solution is of interest. In the higher-order formulation, the tangent terms can be expanded in a series as $\tan x = x + x^3/3 + \ldots$ with respect to the small parameter $p/\lambda$.    In the deep-subwavelength limit, where $p/\lambda \rightarrow 0$, the zero-order approximation (commonly referred to as the quasi-static limit) becomes sufficient. In this regime, the effective refractive indices depend solely on the fractions of the grating materials:
\begin{equation} \label{eq:TE_Eff_med_0}
n_{0,||}^2=(1-F)n_{g}^2+Fn_r^2
\end{equation}
\begin{equation} \label{eq:TM_Eff_med_0}
n_{0,\perp}^2=\left( \frac{(1-F)}{n_{g}^2}+\frac{F}{n_r^2} \right) ^{-1}
\end{equation}
This simple physical model enables the grating to be replaced by an equivalent layer of uniaxial crystal with its optical axes parallel and perpendicular to the grating grooves.   It assumes that the electromagnetic fields cannot resolve the internal structure of the grating and there are no resonances or diffraction effects.   This approximation is valid when the refractive-index contrast between the ridge and groove materials is small \cite{brundrett_subwavelength_1996, lalanne_artificial_2003}, and it is widely used in the design of metal grid polarizers \cite{lalanne_artificial_2003}.

The quasi-static assumption is not valid when the period of the grating becomes comparable to the wavelength ($\lambda/p \sim 1$). In this regime, higher-order terms in the expansion must be taken into account \cite{brundrett_homogeneous_1994, kikuta_ability_1995}.  Rytov analyzed this problem and derived the following second-order approximations for the effective refractive indices: 
\begin{equation} \label{eq:TE_Eff_med_2}
n_{2, ||}^2\approx n_{0, ||}^2+\frac{1}{3} \left[ \pi \frac{p}{\lambda}F(1-F)(n_r^2-n_g^2) \right] ^2,
\end{equation}
\begin{equation} \label{eq:TM_Eff_med_2}
n_{2, \perp}^2\approx n_{0, \perp}^2+\frac{1}{3} \left[ \pi \frac{p}{\lambda}F(1-F) \left( \frac{1}{n_g^2}-\frac{1}{n_r^2} \right) n_{0,\perp}^3n_{0,||} \right] ^2
\end{equation}
For non-dispersive materials, where $n_r$ and $n_g$ are frequency independent, a positive dispersion is expected, since the correction term depends on the even power of $p/\lambda$. These second-order approximations have been widely applied in the design of anti-reflection coatings and waveplates \cite{lalanne_artificial_2003}.  

\subsection{Achromatic condition} \label{achromatic_condition}

An ideal achromatic quarter waveplate should maintain a phase retardation at $\pi/2$ over a broad frequency range.   In this work, we consider a waveplate operating under normal incidence. Owing to the symmetry and periodicity of the grating, only a discrete set of even-symmetric modes with effective refractive indices $n_{\perp}^{(i)}$ and $n_{||}^{(i)}$, where $i=0, 2, 4...$,  can be excited in the grating. The operating regime of the grating is defined by the number of supported modes. Here, we focus on optimizing a grating operating in the single-mode regime, where the design relies on the propagation of a single mode ($i = 0$) and effects associated with modal interference are absent. In the following, the mode index $i = 0$ is omitted for simplicity.

In this regime, the phase retardation between the parallel and perpendicular polarization components must satisfy the following condition over a broad frequency range: 
\begin{equation} \label{eq:refr_achr_cond}
  \Delta \phi = \frac{2\pi \cdot d}{\lambda}(n_{\perp}-n_{||})=\frac{\pi}{2}
\end{equation}
Equivalently, the refractive-index difference must scale linearly with wavelength,:
\begin{equation} \label{eq:eff_med_achr_cond}
  \Delta n = n_{\perp}-n_{||}=\frac{\lambda}{4d} 
\end{equation}
where $\frac{\lambda}{4 d}$  is a slope line to the refractive index difference and the gradient of the line is determined by the grating depth. 

\section{Numerical schemes}

The transcendental equations~\ref{eq:TE_Eff_med} and \ref{eq:TM_Eff_med} were solved using the standard MATLAB (R2021a) function \textit{fsolve}. Initial guesses were selected using an enumeration approach within the range $[0, n_r]$.  Equation~\ref{eq:refr_achr_cond} was then used to calculate the phase retardation. In addition, the effective-medium model was combined with a scattering-matrix formalism \cite{byrnes2016multilayer} to compute the transmission characteristics of the gratings.  

For numerical simulations, we used two approaches. To validate the analytical results, we used frequency domain solver, while for replicating experimental conditions and respective data processing, we used time-domain simulations. 
 
\begin{figure}[htp]
    \centering
    \includegraphics[width=7cm]{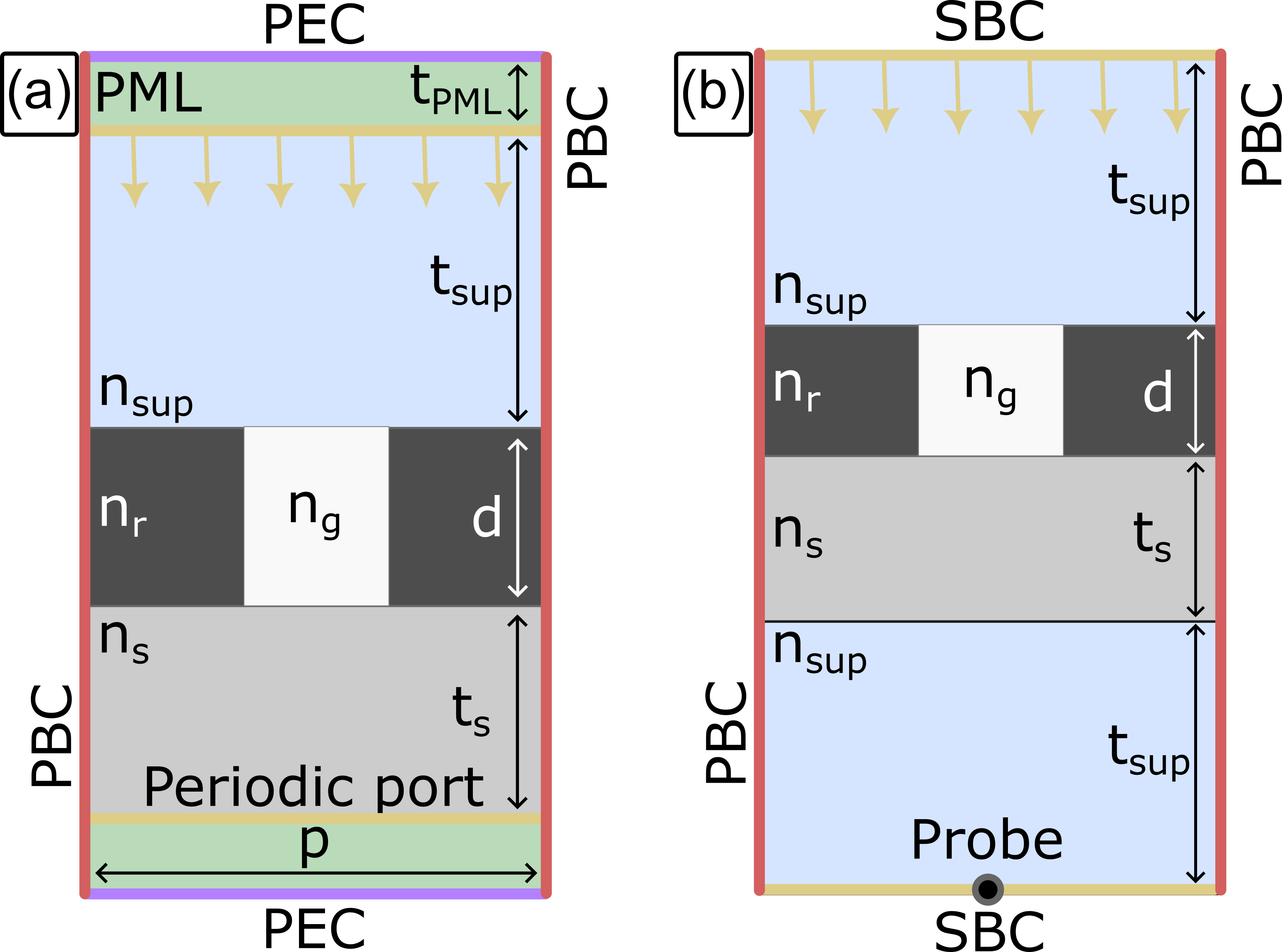}
    \caption{Simulation domains for the COMSOL numerical schemes: frequency domain simulation (a) and time-domain simulation (b).} 
     \label{fig:Figure_2}
\end{figure}

To verify analytical results, we performed numerical simulations using  frequency domain solver of the Radio Frequency (RF) Module of the commercially available COMSOL Multiphysics software (version 6.3) \cite{noauthor_rf_nodate}. The structure under consideration was modeled using a two-dimensional numerical scheme, as shown in Fig.~\ref{fig:Figure_2} (a). Periodic boundary conditions were applied to the left and right sides of the Floquet cell, while perfectly matched layers (PMLs) were placed along the z-direction.  An excitation port was positioned between the PML and the air domains with the slit condition on the interior port. This configuration enabled to track the change in the incident wave and absorb any non-physical or higher-order modes that may be excited. The listener port is placed at the bottom of simulation area, between the substrate domain and the PML domain with the same port condition. The thicknesses of both the air and substrate domains were chosen to be sufficiently large to prevent interactions between the grating modes and the ports. The substrate was modeled as semi-infinite in the z-direction to avoid Fabry-Perot resonances which can arise in the substrate of finite thickness. The meshing was set to the “extremely fine”, and it was built automatically. To analyze the grating responses to the incident beam,  we conducted separate simulations  for perpendicular and parallel polarizations. Based on the computed complex transmission amplitudes, we reconstructed the phase retardation as
\begin{equation}
    \Delta\varphi=\arg(t_{\perp}\cdot t_{||}^*)
\end{equation}
where $^*$ denotes  the complex conjunction, $t_{\perp}$ and $t_{||}$  are the complex transmission amplitudes for the perpendicular and parallel polarizations, respectively. 

To verify experimental results, rigorous numerical simulations were performed using  transient solver of the RF Module \cite{noauthor_rf_nodate}. A two-dimensional numerical scheme, shown in Fig.~\ref{fig:Figure_2}(b), was employed. Periodic boundary conditions were applied to the left and right sides of the Floquet cell consisting of one grating period, while scattering boundary conditions (SBCs) were placed along the z-direction. The upper SBC was used to excite a pulse with a time-dependent electric field derived from experimental measurements (reference pulse). A probe located at the center of the lower SBC recorded the transmitted pulse. Frequency-domain spectra containing both amplitude and phase information were obtained by applying a fast Fourier transform (FFT) to the time-domain signal.  The phase response was reconstructed using the phase-retrieval algorithm described in\cite{jepsen2019phase}. To analyze the grating response under different excitation conditions, separate simulations were carried out for parallel and perpendicular polarizations.   The mesh was configured to satisfy a Courant–Friedrichs–Lewy number of 0.06.  In all simulations, the grating ridges were modeled as silicon with a nondispersive refractive index of $n_s=n_r=3.416$, while the grating grooves and surrounding media were assumed to be vacuum with $n_{sup}=n_g=1$.

\section{Experimental setup}
\begin{figure}[h!]
    \centering
    \includegraphics[width=8cm]{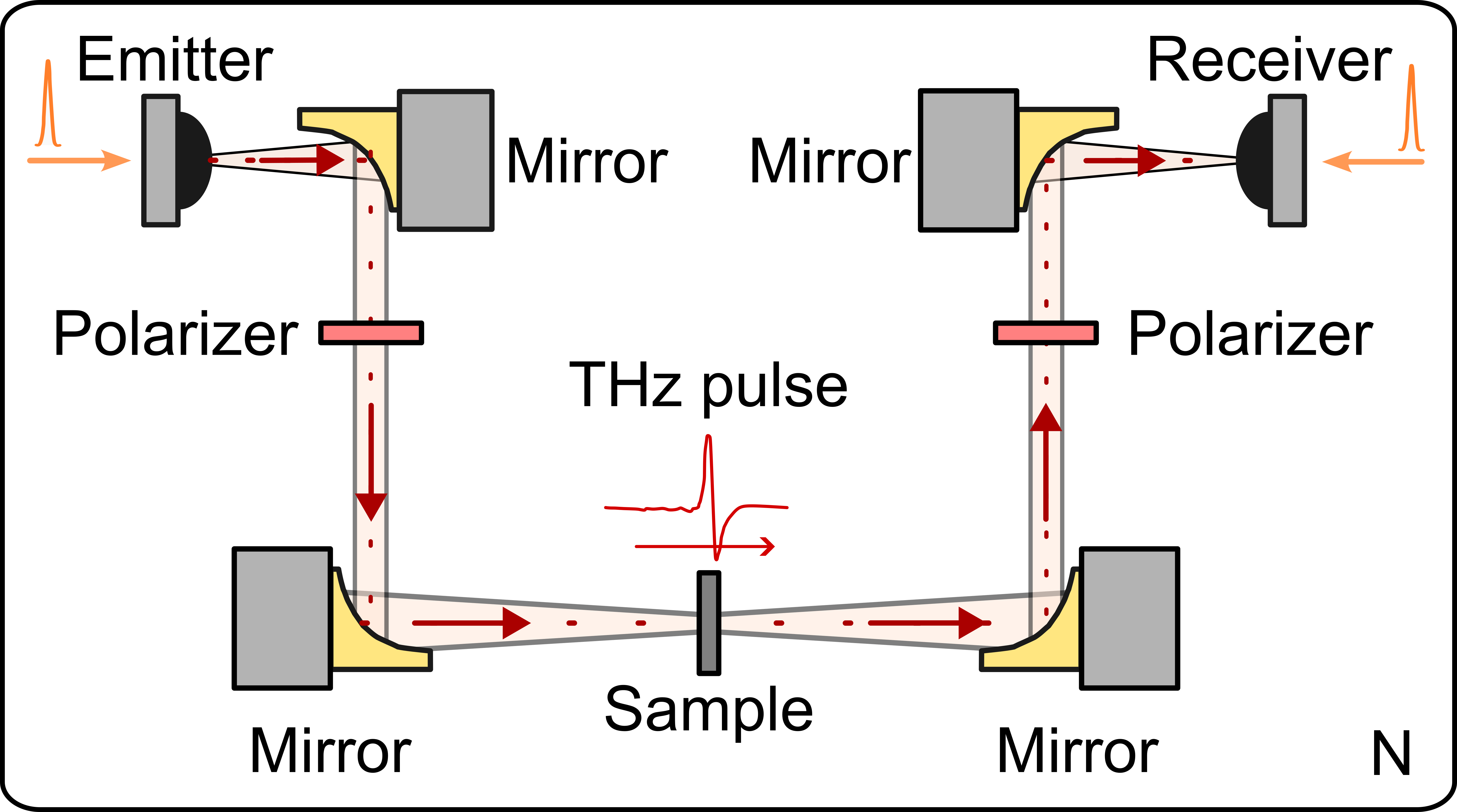}
    \caption{Schematics of the Terahertz part of the time-domain spectrometer.}
    \label{fig:Figure_3}
\end{figure}

The fabricated gratings were studied experimentally using an EKSPLA terahertz time-domain spectrometer.
The system is equipped with a CALMAR Mendocino fiber laser (FPL-04RCFF), delivering pulses with a duration $<90$ fs and a repetition rate of 100 MHz with a center wavelength of 780 nm. The setup is enclosed in a nitrogen-purged atmosphere to prevent water-vapor absorption of the THz radiation.  The "terahertz part" of the setup is shown in Fig.~\ref{fig:Figure_3}. 
The terahertz pulse from the emitter was directed onto a sample using a pair of parabolic mirrors. After transmission through the sample, the beam was collected and guided to the receiver by another pair of parabolic mirrors. Two polarizers were used to purify the polarization and define the polarization axis of the system. The sample was positioned at the common focus of the parabolic mirrors to ensure the validity of the plane-wave approximation. Three time-domain waveforms were recorded: a reference signal corresponding to transmission through a 10 mm diaphragm, and two signals corresponding to the orientations where the grating grooves were parallel and perpendicular to the polarization of the incident beam. These positions were  determined experimentally based on maximum and minimum of the amplitude of the detected waveforms. Frequency-domain spectra were obtained in the same manner as described in the "Numerical simulations" section.
 
\section{Fabrication}

\begin{figure}[h!]
    \centering
    \includegraphics[width=6cm]{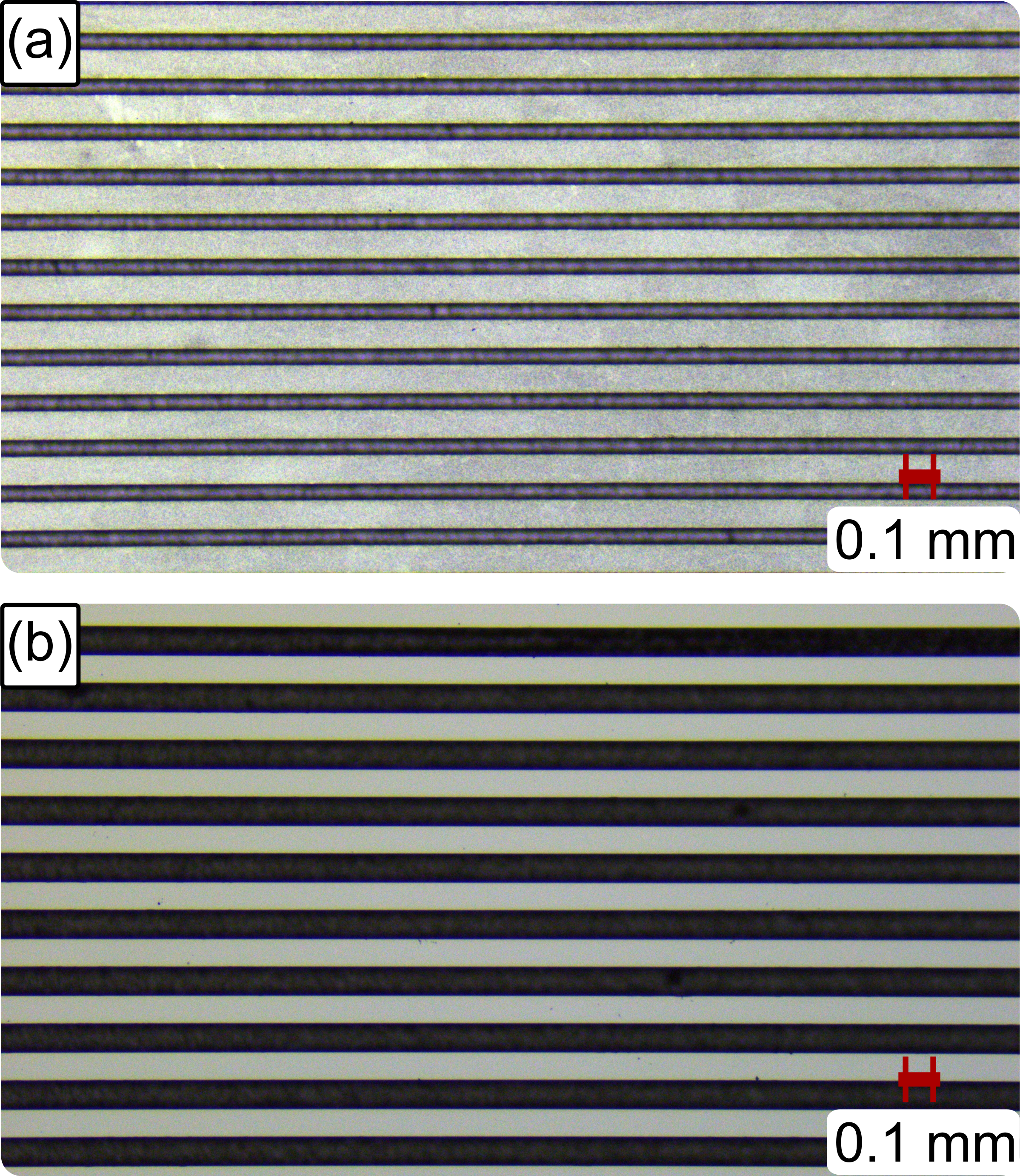}
    \caption{Microscope images of the fabricated gratings: (a) Sample I: p=$105\pm1 \mu m$, w=$38\pm1 \mu$m, $129\pm1 \mu$m  and (b) Sample II: p=$132 \pm 1 \mu$m, w=$69\pm1 \mu$m, $119\pm1 \mu$m.}
    \label{fig:Figure_4}
\end{figure}

As a proof-of-concept, we fabricated two rectangular dielectric gratings. The first grating (Sample I) had a period of $105\pm1 \mu m$, a groove width of $38\pm1 \mu$m and a depth of $129\pm1 \mu$m (Fig.~\ref{fig:Figure_4}(a)). The second grating (Sample II) had a period of $132 \pm 1 \mu$m, a groove width of $69\pm1 \mu$m and a depth of $119\pm1 \mu$m (Fig.~\ref{fig:Figure_4}(b)). The gratings, each with an area of 10x10 mm$^2$, were fabricated on a high resistivity floating zone silicon (HRFZ-Si) wafer with an overall thickness of 525 $\mu$m.

A high-resistivity float-zone (HRFZ) silicon wafer was sequentially cleaned using acetone, isopropyl alcohol (IPA), and deionised (DI) water. The wafer was subsequently dehydrated on a hotplate at 120 °C for 120 sec to remove residual moisture. AZ 5214E photoresist was spin-coated at 3000 rpm for 30 s, followed by a soft bake at 95 °C for 60 s.

Photolithographic patterning was carried out using an MLA150 maskless aligner (Heidelberg Instruments). After exposure, the resist was developed in a metal-ion-free (MIF) developer to reveal the patterned features. A 120 nm-thick chromium (Cr) layer was then deposited and used as a hard mask.

Subsequently, silicon dry etching was performed using a standard Bosch process in a deep reactive ion etching (DRIE) system (Oxford Instruments PlasmaPro Estrelas). After achieving the desired etch depth, the Cr hard mask was removed via a wet etching process using a chromium etchant.

 \section{Results}

\subsection{Optimization}

Direct optimization of the grating by searching for the ideal achievable dependence of the index difference on wavelength required by Eq.~\ref{eq:eff_med_achr_cond} is not convenient. To formalize the optimization procedure and simplify the search for achromatic operation, we analyze the dependence of the optimal grating depth $d_{opt}$ on frequency $f=c/\lambda$, where $c$ is the speed of light:
\begin{equation}
d_{opt} = \frac{\lambda \cdot \pi/2}{2\pi \cdot \Delta n} = \frac{c}{4f \Delta n}
\end{equation}
This expression reformulates the achromatic condition (Eq.~\ref{eq:eff_med_achr_cond}) and enables direct determination of the grating depth required to produce a $\pi/2$ phase difference for two orthogonal polarizations. The frequency dependence of $d_{opt}$ reflects the phase response of the structure for a given set of grating parameters, namely the period $p$ and the grooves width $w$. In the ideal case, $d_{opt}$ should be constant for a wide frequency range, and we are exploring how close we can get to this condition.

\begin{figure}[t]
    \centering
    \includegraphics[width=7cm]{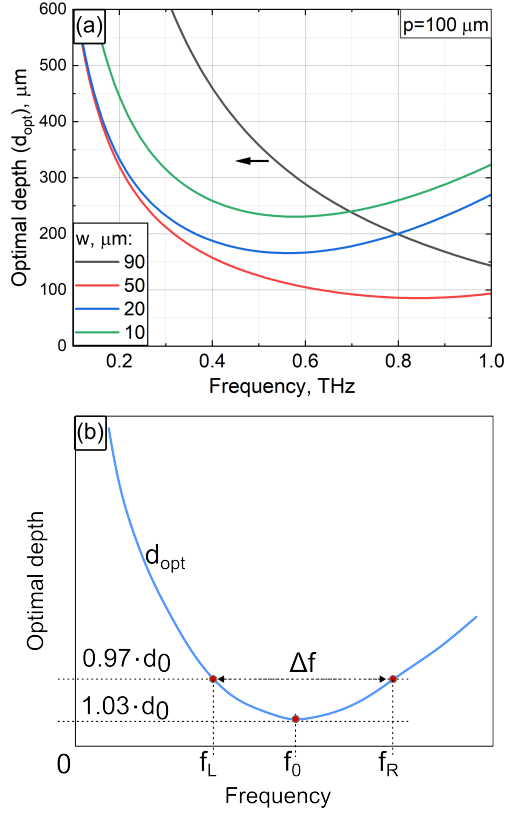}
    \caption{ (a) Optimal depth of Si grating required to achieve $\pi$/2 phase retardation, for $p=100 \mu m$ and various groove widths; (b) Schematic illustration of parameters for the optimization algorithm.}
    \label{fig:Figure_5}
\end{figure}

To illustrate the optimization strategy, we examine the behavior of $d_{opt}$ for different grating geometries over the frequency range from 0.1 to 1 THz. As shown in Fig.~\ref{fig:Figure_5}(a), under specific fill factors the dependence of  $d_{opt}$ on frequency exhibits a parabolic-like profile.  Since equations~\ref{eq:TE_Eff_med} and \ref{eq:TM_Eff_med} do not contain physical constants, this behavior is scale-invariant.

Nearly achromatic operation is achieved  when this parabolic-like curve exhibits minimal variation around its dip, corresponding to a central frequency $f_0$ and  associated depth $d_0$ (see Fig.\ref{fig:Figure_5}(b)).  We define the acceptable deviation of the phase retardation within $3 \%$ from $\frac{\pi}{2}$. Given the linear dependence of the phase retardation on the groove depth   (Eq.~\ref{eq:eff_med_achr_cond}), the final optimized depth is chosen as $d_{opt} \cdot 1.03$. Within this definition, the waveplate bandwidth $\Delta f=f_R-f_L$ is determined by the intersections of the parabolic-like $d_0(f)$ curve with the $0.97 \cdot d_0$ line. Maximizing this bandwidth around a target frequency is the primary optimization goal which can be achieved by fixing the target frequency and varying the groove period and width.
\subsection{Bandwidth}

We first applied the proposed optimization procedure to determine the optimal grating parameters for operation at a target frequency of $f_0=0.6$~THz. The groove width was varied from $0.1 p$ to $0.9 p$. In this optimization, the modes interference and Fabry-Perot were neglected as we used only equations \ref{eq:TE_Eff_med} and \ref{eq:TM_Eff_med}. This approach allows us to capture the ideal phase response of the grating. 

\begin{figure}[!h]
    \centering 
    \includegraphics[width=8cm]{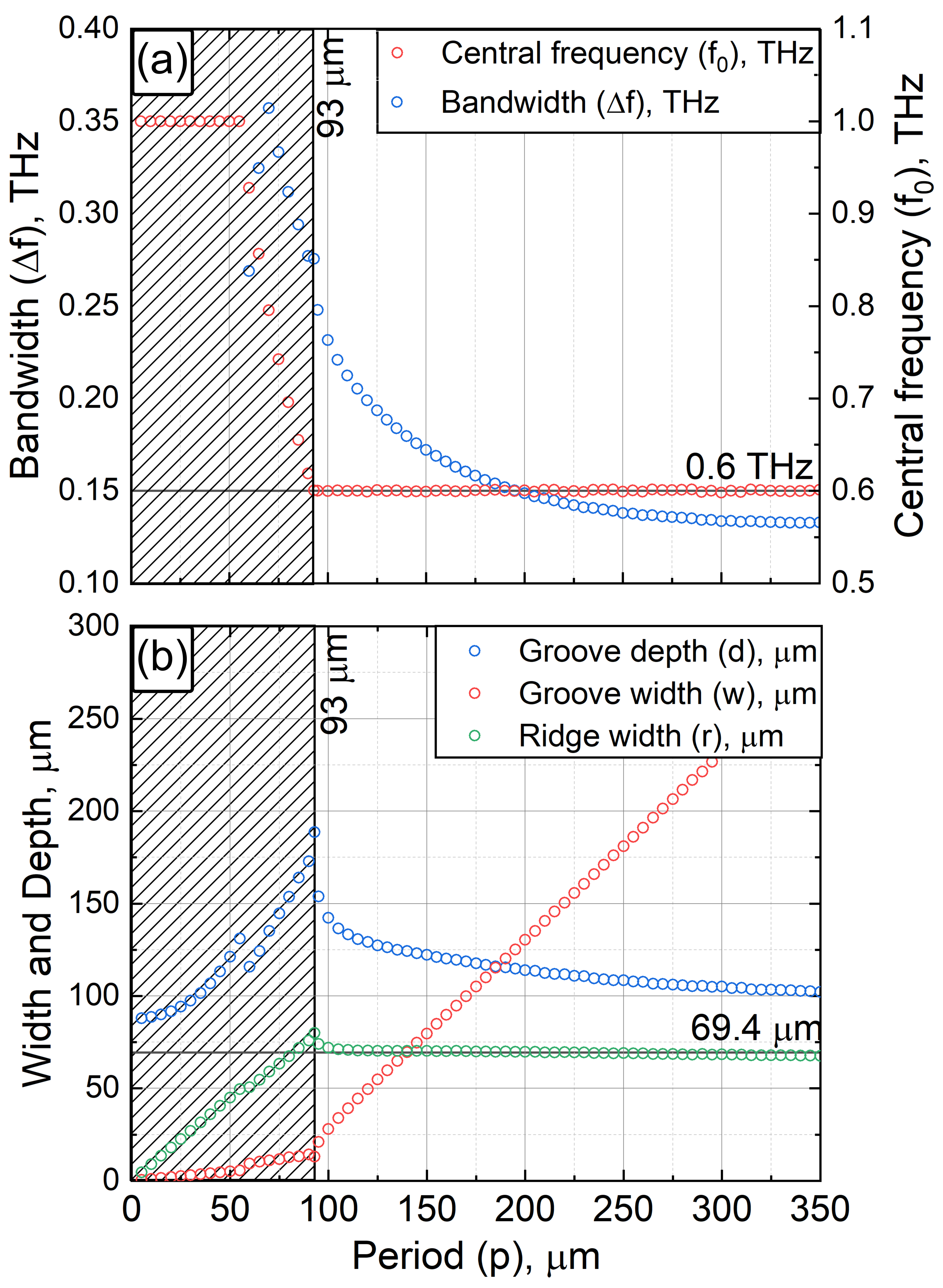}
    \caption{
    (a) Dependence of the optimized silicon grating bandwidth and central frequency on structure period; (b) Corresponding optimal geometric parameters required to achieve $\pi/2$ phase retardation. In both panels, shading shows the region of parameters, where the method was not able to find required band with the central frequency at 0.6 THz.}
    \label{fig:Figure_6}
\end{figure}

Fig.~\ref{fig:Figure_6}(a) shows  the dependence of the optimized bandwidth and central frequency as a function of the grating period. Below the value of the period of approximately 93 $\mu m$, the method was not able to identify appropriate phase behavior for the waveplate at a target frequency of 0.6 THz. We also note that the bandwidth decays for larger periods.  

Fig.~\ref{fig:Figure_6}(b)  presents the optimized groove width,  ridge width, and groove depth. The large bandwidth is associated with deeper grooves and narrower groove width. The ridge width remains nearly constant, with an average value of approximately $69.4$ $\mu m$, while the groove width increases linearly with the period. In contrast, the optimal groove depth decreases as the period increases. These trends will define a trade-off between fabrication constraints and the maximum achievable bandwidth, since fabrication methods have limitation on the aspect ratio of etched grooves.

This observed behavior, when the phase difference exhibits the parabolic-like dependence, is resonant in nature. The ridge width determines the resonance wavelength associated with the excitation of the second transverse magnetic Bloch mode in the structure~\cite{liu_highly_2019}: 

\begin{equation} \label{eq:magnetic_dipole}
    \lambda_0\approx2\cdot n_rr
\end{equation}

For example, at the target frequency of 0.6 THz ($\lambda_0=499.7 \mu m$), the optimal ridge width is $r \approx 73~\mu$m. Under this condition, scattering is maximized and the effective refractive index difference becomes proportional to the wavelength. This observation significantly simplifies the optimization procedure described above.   

\begin{figure}[!h]
    \centering
    \includegraphics[width=9cm]{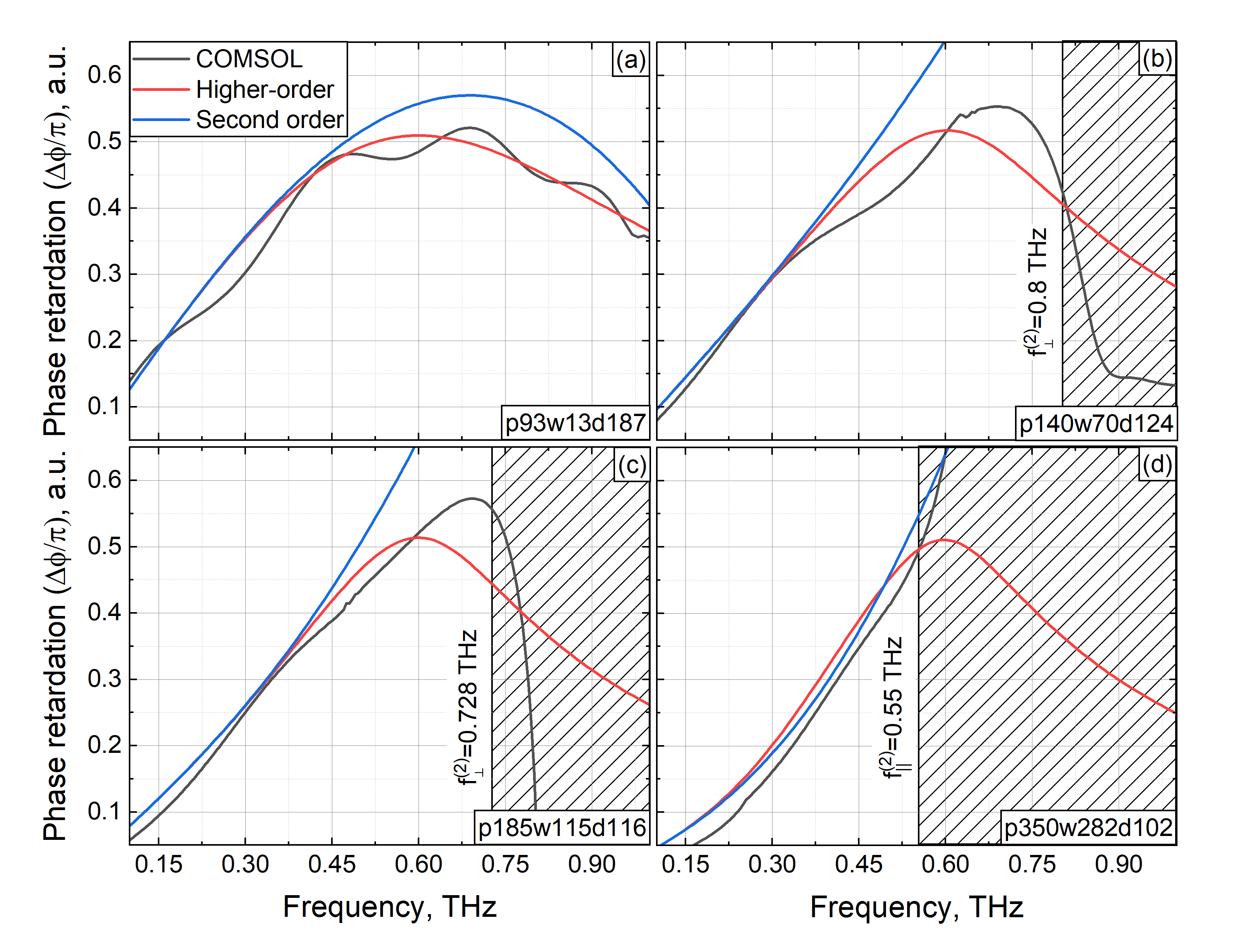}
    \caption{Comparison of the effective medium approximations with full wave numerical simulations for different grating parameters. (a) p = 93 $\mu m$, w = 13 $\mu m$, d = 187 $\mu m$; (b) p = 140 $\mu m$, w = 70 $\mu m$, d = 124 $\mu m$; (c) p = 185 $\mu m$, w = 115 $\mu m$, d = 116 $\mu m$; (d) p = 350 $\mu m$, w = 283 $\mu m$, d = 102 $\mu m$. Shaded area shows the regions where multiple modes are excited in the grating.}
    \label{fig:Figure_7}
\end{figure}

The next step is to compare the phase retardation obtained from the analytical model (see Eq.~\ref{eq:refr_achr_cond}) with rigorous full-wave frequency-domain simulations. Several grating configurations were selected for this comparison: a design exhibiting the maximum bandwidth (Fig.~\ref{fig:Figure_7}(a)), a structure with a period-to-width ratio of $p/w=0.5$ (Fig.~\ref{fig:Figure_7}(b)), a configuration with equal groove width and depth $d=w$ (Fig.~\ref{fig:Figure_7} (c)) and a structure corresponding to the minimum bandwidth among the simulated cases (Fig.~\ref{fig:Figure_7}(d)).

Fig.~\ref{fig:Figure_7} (a) demonstrates that the higher-order effective medium theory correctly predicts both the overall phase distribution and the position of the parabolic-like dependence.  However, Fabry-Perot resonances inside ridges, which are not taken into account in the effective medium theory,  lead to distortions of the phase response. According to the second-order effective medium theory, such deviations are expected to emerge at higher frequencies, whereas the quasi-static regime remains valid at lower frequencies, and only qualitatively correctly describes the phase retardation in the whole considred frequency range.

For larger grating periods, the theoretical model becomes less accurate. As shown in Fig.~\ref{fig:Figure_7} (b,c), the phase retardation still exhibits a parabolic-like profile, but the central frequency is noticeably shifted. This shift arises from the excitation of higher-order modes at approximately 0.8 THz and 0.728 THz, respectively, this regions are shaded. These modes modify the overall phase response through modal interference, an effect that is not captured by the single-mode analytical model. In the case of the grating with a period of 350$\mu m$, a higher-order mode is already supported at 0.55 THz, indicating that the single-mode regime is no longer valid. Consequently, the predictions of the higher-order effective medium theory break down for this configuration.  
 Considering that the higher-order effective medium theoretical description is valid at the onset of the parabolic-like regime corresponding to the maximum achievable bandwidth, we estimate the relative bandwidth applicable at other THz frequencies for a given geometry and material parameters as  $\Delta f /f_0 \cdot 100\% \approx46 \%$.
  
 In these simulations, the groove width was constrained to the range $0.1p$ and  $0.9p$. Consequently, the obtained bandwidth does not represent the theoretical maximum, but takes into account reasonable practical limitations. Nevertheless, even within these constraints, the optimized geometries require extremely large aspect ratios (ration of the depth of the grooves to their width), posing significant fabrication challenges. For example,  at $f_0=0.6$ THz,  the optimized structure yields a width-to-depth ratio of $w:d\approx13:187\approx1:14$,  a value which is difficult to achieve using standard microfabrication techniques.
 
\subsection{Experimental results}

\begin{figure}[h!]
    \centering
     \includegraphics[width=8cm]{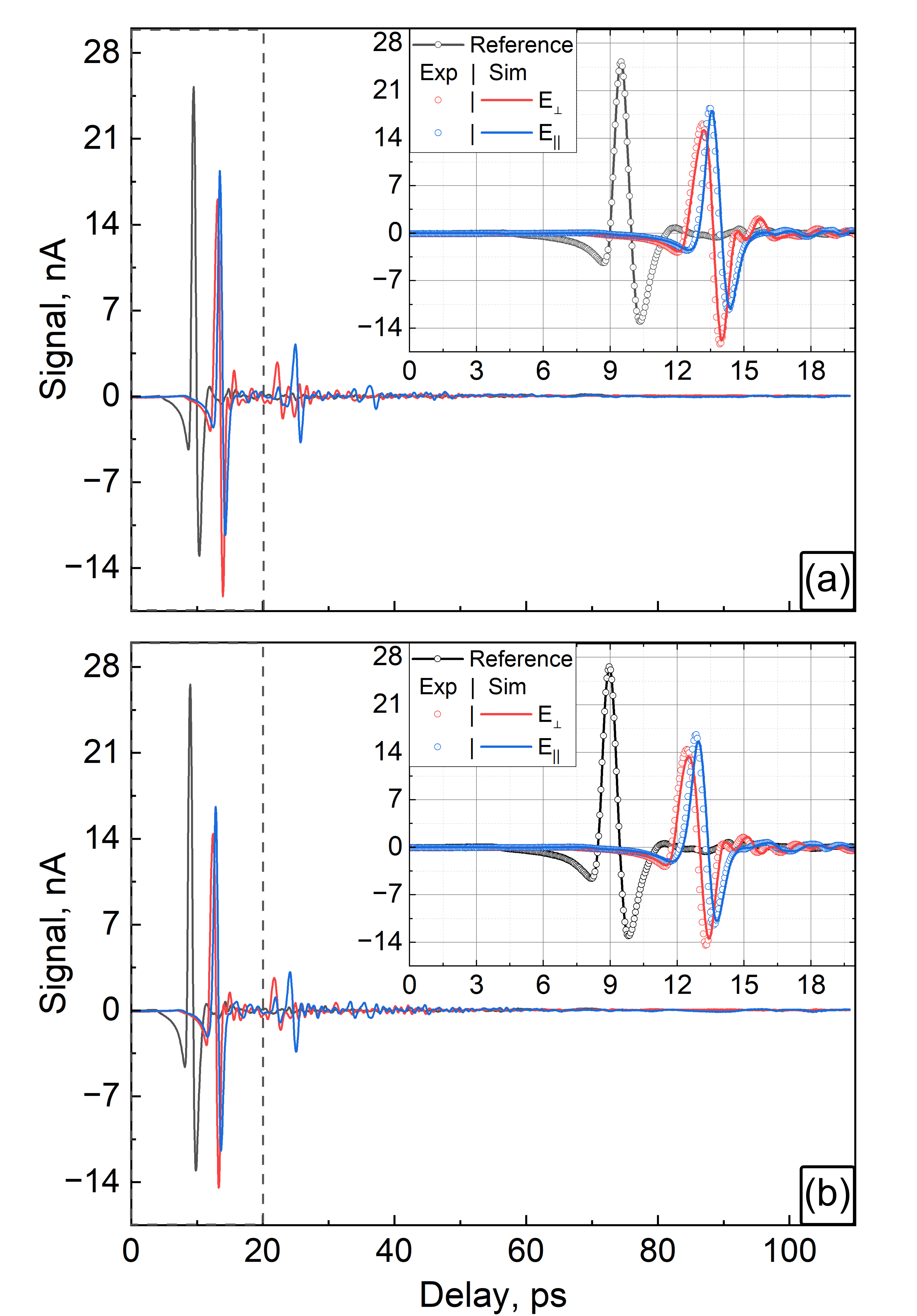}
    \caption{Time domain traces of the reference signal and the two transmitted signals for different polarizations. (a) Sample I and (b) Sample II. In numerical simulations, the reference signal was used as an incident pulse, and the simulated transmitted pulses of both polarizations match well the measurements.}
    \label{fig:Figure_8}
\end{figure}

\begin{figure}[!h]
    \centering
     \includegraphics[width=8cm]{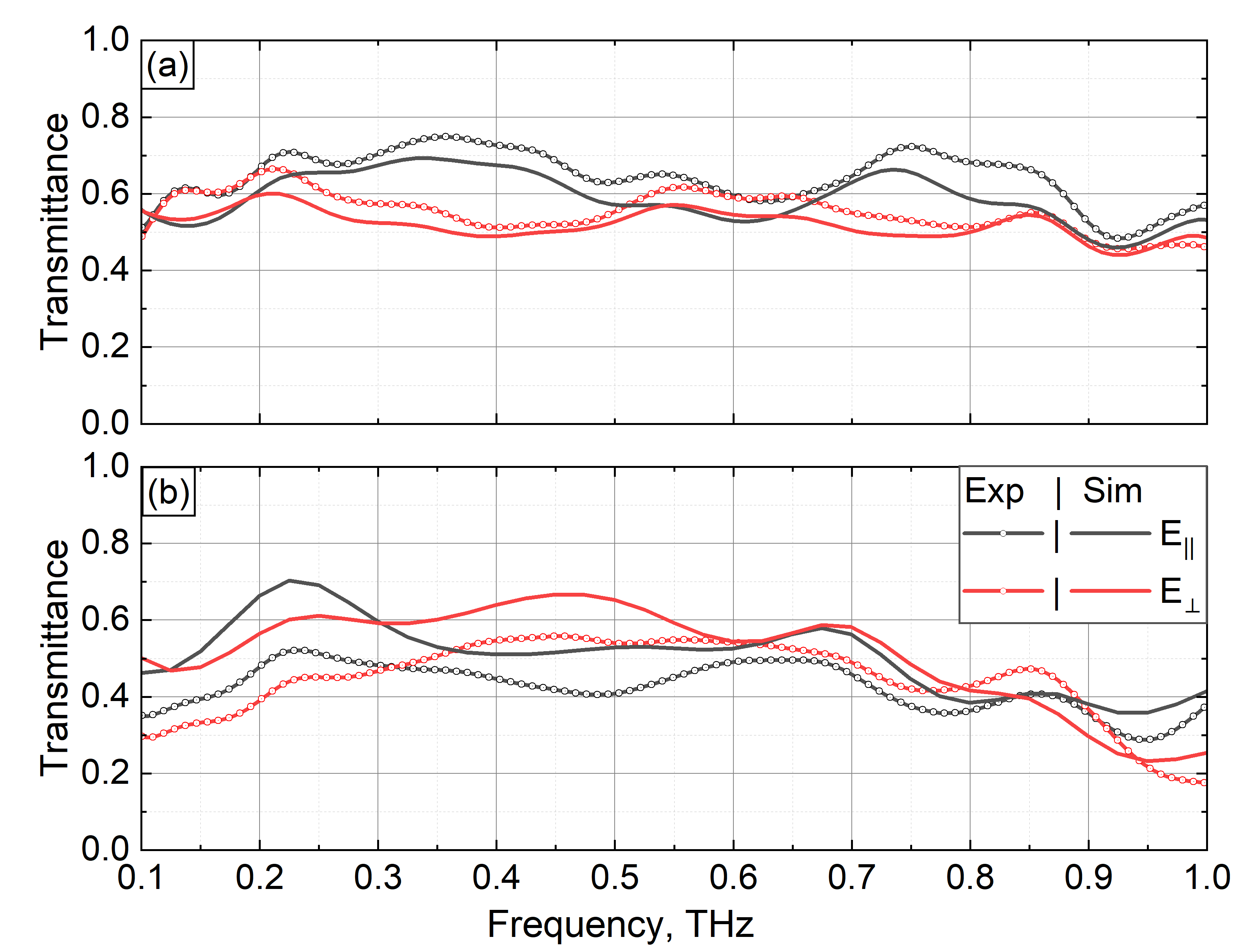}
    \caption{Comparison of the simulated and experimentally measured transmission spectra corresponding to  (a) Sample I and (b) Sample II.}
    \label{fig:Figure_9}
\end{figure}

\begin{figure}[!h]
    \centering
     \includegraphics[width=8cm]{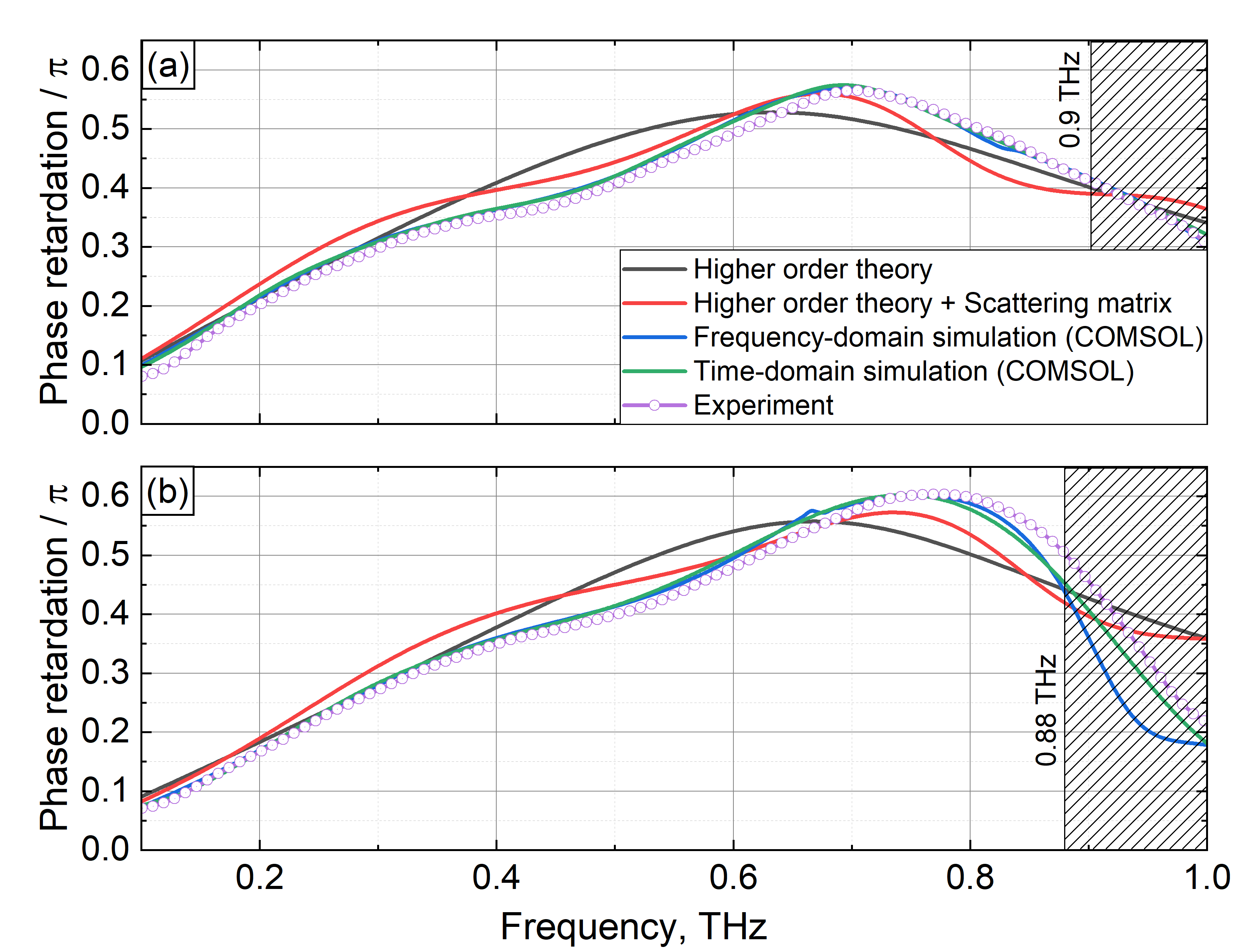}
    \caption{Experimentally measured, theoretically calculated and numerically simulated phase retardation for (a) Sample I and (b) Sample II. Shaded area shows the regions where multiple modes are excited in the grating. }
    \label{fig:Figure_10}
\end{figure}

 The experimental and simulated time-domain, transmission spectra  for the perpendicular ($E_\perp$) and parallel ($E_{||}$) polarization components, as well as the corresponding phase retardation for Sample I and Sample II, are shown in Figs. \ref{fig:Figure_8}-\ref{fig:Figure_10}.

The time-domain signals contain multiple weak echo contributions originating from Fabry–Pérot resonances within the substrate. To suppress their influence on the final spectra, a rectangular temporal window was applied to both the experimental and simulated data. A remaining temporal window of 20~ps was sufficient to accurately reproduce the grating response, as no narrow resonances requiring higher spectral resolution are present.

The simulated time-domain signals are in a good agreement with the experimental results. This agreement confirms that the lateral dimensions of the samples are sufficiently large to be treated as infinite gratings in the simulations, and it validates the accuracy of the geometrical parameters as well as the assumption of normal incidence in the experiment.

Despite the removal of Fabry–Perot resonances originating from the substrate, the transmission spectra remain distorted due to Fabry–Perot resonances within the grating cavities, which cannot be fully eliminated using this approach (see Fig.~\ref{fig:Figure_9}). An ideal waveplate should exhibit equal transmission for both polarizations; therefore, additional strategies are required to suppress this effect. The relatively low overall transmission is primarily attributed to strong reflections at the substrate–air interface. Furthermore, the measured spectra are influenced by the non-ideal spatial profile of the incident beam and its alignment.

Fig.~\ref{fig:Figure_10} shows the phase retardation across the investigated frequency range. The phase retardation was obtained using five different methods, reflecting various levels of approximation. The simplified higher-order effective-medium analysis (black curve) neglects Fabry–Perot effects and predicts an ideal, undistorted parabolic-like phase-retardation behavior within the single-mode regime (< 0.9 THz). The scattering-matrix approach (red curve), assuming an infinite substrate ($n_s=n_r$), captures the influence of Fabry–Perot resonances within the grating ridges. The phase obtained using the rigorous frequency-domain method (blue curve) described in the Numerical schemes section accounts for additional interactions within the grating that are not captured by effective-medium theory. The time-domain method further incorporates the non-ideal nature of the pulsed excitation and experimental system characteristics and frequency filtering (green curve). The experimental results (purple curve) reproduce the overall phase-retardation behavior predicted by theory and numerical simulations. Minor deviations are attributed to the high sensitivity of the phase response to geometrical tolerances and beam alignment.

\section{Conclusion}

We have investigated rectangular high-contrast dielectric gratings as quarter-wave plates operating in the terahertz range. A systematic design strategy and theoretical framework were developed to analyze the phase response and bandwidth limitations of such structures. We showed that the theoretical upper limit of the achievable relative bandwidth in the single-mode regime is approximately $46\%$. Further bandwidth enhancement is fundamentally constrained by the intrinsic properties of single-layer rectangular dielectric gratings and by practical fabrication limitations. 

These constraints arise primarily from Fabry-Perot resonances within the substrate and grating ridges, as well as from the higher-order mode excitation. Suppressing these effects could enable the realization of highly efficient waveplates based on this design although with a limited operational bandwidth.  To achieve broader bandwidths, alternative strategies must be considered, including multilayer or stacked grating designs, groove-profile optimization, or the use of dispersive materials in place of non-dispersive dielectrics. 

\section*{Acknowledgments}
This work was supported by the Australian Research Council under the Centers of Excellence program (CE200100010).

This work was performed in part at the Melbourne Centre for Nanofabrication (MCN) in the Victorian Node of the Australian National Fabrication Facility (ANFF).

\bibliographystyle{unsrt}  
\bibliography{Bibliography.bib}

@article{masson_terahertz_2006,
	title = {Terahertz achromatic quarter-wave plate},
	volume = {31},
	copyright = {https://doi.org/10.1364/OA\_License\_v1\#VOR},
	issn = {0146-9592, 1539-4794},
	url = {https://opg.optica.org/abstract.cfm?URI=ol-31-2-265},
	doi = {10.1364/ol.31.000265},
	language = {en},
	number = {2},
	urldate = {2025-07-25},
	journal = {Optics Letters},
	author = {Masson, Jean-Baptiste and Gallot, Guilhem},
	month = jan,
	year = {2006},
	note = {Publisher: Optica Publishing Group},
	pages = {265},
}

@article{chen_terahertz_2013,
	title = {Terahertz achromatic quarter wave plate: {Design}, fabrication, and characterization},
	volume = {311},
	issn = {0030-4018},
	shorttitle = {Terahertz achromatic quarter wave plate},
	url = {https://linkinghub.elsevier.com/retrieve/pii/S0030401813007694},
	doi = {10.1016/j.optcom.2013.08.039},
	language = {en},
	urldate = {2025-07-25},
	journal = {Optics Communications},
	author = {Chen, Zaichun and Gong, Yandong and Dong, Hui and Notake, Takashi and Minamide, Hiroaki},
	month = jan,
	year = {2013},
	note = {Publisher: Elsevier BV},
	pages = {1--5},
}

@article{zhang_general_2023,
	title = {General approach of terahertz achromatic quarter-wave plate composed of stacked quartz plates},
	volume = {2},
	copyright = {https://doi.org/10.1364/OA\_License\_v2\#VOR-OA},
	issn = {2770-0208},
	url = {https://opg.optica.org/abstract.cfm?URI=optcon-2-7-1597},
	doi = {10.1364/optcon.494818},
	language = {en},
	number = {7},
	urldate = {2025-07-25},
	journal = {Optics Continuum},
	author = {Zhang, Tianmiao and Kropotov, Grigory and Khodzitsky, Mikhail},
	month = jul,
	year = {2023},
	note = {Publisher: Optica Publishing Group},
	pages = {1597},
}

@inproceedings{zhang_terahertz_2021,
	address = {Chengdu, China},
	title = {Terahertz achromatic quarter-wave plate composed of quartz and {MgF}$_{\textrm{2}}$},
	copyright = {https://doi.org/10.15223/policy-029},
	url = {https://ieeexplore.ieee.org/document/9567029/},
	doi = {10.1109/irmmw-thz50926.2021.9567029},
	language = {en},
	urldate = {2025-07-25},
	booktitle = {2021 46th {International} {Conference} on {Infrared}, {Millimeter} and {Terahertz} {Waves} ({IRMMW}-{THz})},
	publisher = {IEEE},
	author = {Zhang, T. and Popov, D. and Khodzitsky, M.},
	month = aug,
	year = {2021},
	pages = {1--2},
}

@article{withayachumnankul_fundamentals_2014,
	title = {Fundamentals of {Measurement} in {Terahertz} {Time}-{Domain} {Spectroscopy}},
	volume = {35},
	copyright = {http://www.springer.com/tdm},
	issn = {1866-6892, 1866-6906},
	url = {http://link.springer.com/10.1007/s10762-013-0042-z},
	doi = {10.1007/s10762-013-0042-z},
	language = {en},
	number = {8},
	urldate = {2025-07-25},
	journal = {Journal of Infrared, Millimeter, and Terahertz Waves},
	author = {Withayachumnankul, Withawat and Naftaly, Mira},
	month = aug,
	year = {2014},
	note = {Publisher: Springer Science and Business Media LLC},
	pages = {610--637},
}

@article{neu_tutorial_2018,
	title = {Tutorial: {An} introduction to terahertz time domain spectroscopy ({THz}-{TDS})},
	volume = {124},
	issn = {0021-8979, 1089-7550},
	shorttitle = {Tutorial},
	url = {https://pubs.aip.org/jap/article/124/23/231101/155935/Tutorial-An-introduction-to-terahertz-time-domain},
	doi = {10.1063/1.5047659},
	language = {en},
	number = {23},
	urldate = {2025-07-25},
	journal = {Journal of Applied Physics},
	author = {Neu, Jens and Schmuttenmaer, Charles A.},
	month = dec,
	year = {2018},
	note = {Publisher: AIP Publishing},
}

@article{koch_terahertz_2023,
	title = {Terahertz time-domain spectroscopy},
	volume = {3},
	copyright = {https://www.springernature.com/gp/researchers/text-and-data-mining},
	issn = {2662-8449},
	url = {https://www.nature.com/articles/s43586-023-00232-z},
	doi = {10.1038/s43586-023-00232-z},
	language = {en},
	number = {1},
	urldate = {2025-07-25},
	journal = {Nature Reviews Methods Primers},
	author = {Koch, Martin and Mittleman, Daniel M. and Ornik, Jan and Castro-Camus, Enrique},
	month = jun,
	year = {2023},
	note = {Publisher: Springer Science and Business Media LLC},
}

@article{cong_highly_2014,
	title = {Highly flexible broadband terahertz metamaterial quarter‐wave plate},
	volume = {8},
	copyright = {http://onlinelibrary.wiley.com/termsAndConditions\#vor},
	issn = {1863-8880, 1863-8899},
	url = {https://onlinelibrary.wiley.com/doi/10.1002/lpor.201300205},
	doi = {10.1002/lpor.201300205},
	language = {en},
	number = {4},
	urldate = {2025-07-25},
	journal = {Laser \& Photonics Reviews},
	author = {Cong, Longqing and Xu, Ningning and Gu, Jianqiang and Singh, Ranjan and Han, Jiaguang and Zhang, Weili},
	month = jul,
	year = {2014},
	note = {Publisher: Wiley},
	pages = {626--632},
}

@article{kawada_achromatic_2014,
	title = {Achromatic prism-type wave plate for broadband terahertz pulses},
	volume = {39},
	copyright = {https://doi.org/10.1364/OA\_License\_v1\#VOR},
	issn = {0146-9592, 1539-4794},
	url = {https://opg.optica.org/abstract.cfm?URI=ol-39-9-2794},
	doi = {10.1364/ol.39.002794},
	language = {en},
	number = {9},
	urldate = {2025-07-25},
	journal = {Optics Letters},
	author = {Kawada, Yoichi and Yasuda, Takashi and Nakanishi, Atsushi and Akiyama, Koichiro and Hakamata, Kento and Takahashi, Hironori},
	month = may,
	year = {2014},
	note = {Publisher: Optica Publishing Group},
	pages = {2794},
}

@article{wang_broadband_2015,
	title = {Broadband tunable liquid crystal terahertz waveplates driven with porous graphene electrodes},
	volume = {4},
	copyright = {https://creativecommons.org/licenses/by/3.0/},
	issn = {2047-7538},
	url = {https://www.nature.com/articles/lsa201526},
	doi = {10.1038/lsa.2015.26},
	language = {en},
	number = {2},
	urldate = {2025-07-25},
	journal = {Light: Science \& Applications},
	author = {Wang, Lei and Lin, Xiao-Wen and Hu, Wei and Shao, Guang-Hao and Chen, Peng and Liang, Lan-Ju and Jin, Biao-Bing and Wu, Pei-Heng and Qian, Hao and Lu, Yi-Nong and Liang, Xiao and Zheng, Zhi-Gang and Lu, Yan-Qing},
	month = feb,
	year = {2015},
	note = {Publisher: Springer Science and Business Media LLC},
	pages = {e253--e253},
}

@article{chen_highly_2019,
	title = {Highly {Efficient} {Ultra}‐{Broadband} {Terahertz} {Modulation} {Using} {Bidirectional} {Switching} of {Liquid} {Crystals}},
	volume = {7},
	copyright = {http://onlinelibrary.wiley.com/termsAndConditions\#vor},
	issn = {2195-1071, 2195-1071},
	url = {https://onlinelibrary.wiley.com/doi/10.1002/adom.201901321},
	doi = {10.1002/adom.201901321},
	language = {en},
	number = {24},
	urldate = {2025-07-25},
	journal = {Advanced Optical Materials},
	author = {Chen, Xuequan and Li, Kaidi and Zhang, Rui and Gupta, Swadesh Kumar and Srivastava, Abhishek Kumar and Pickwell‐MacPherson, Emma},
	month = dec,
	year = {2019},
	note = {Publisher: Wiley},
}

@article{zhang_tunable_2020,
	title = {Tunable terahertz phase shifter based on dielectric artificial birefringence grating filled with polymer dispersed liquid crystal},
	volume = {10},
	copyright = {https://doi.org/10.1364/OA\_License\_v1\#VOR-OA},
	issn = {2159-3930},
	url = {https://opg.optica.org/abstract.cfm?URI=ome-10-2-282},
	doi = {10.1364/ome.383058},
	language = {en},
	number = {2},
	urldate = {2025-07-25},
	journal = {Optical Materials Express},
	author = {Zhang, Xin and Fan, Fei and Zhang, Chun-Yue and Ji, Yun-Yun and Wang, Xiang-Hui and Chang, Sheng-Jiang},
	month = feb,
	year = {2020},
	note = {Publisher: Optica Publishing Group},
	pages = {282},
}

@article{petrov_design_2022,
	title = {Design of broadband terahertz vector and vortex beams: {I}. {Review} of materials and components},
	volume = {3},
	issn = {2831-4093},
	shorttitle = {Design of broadband terahertz vector and vortex beams},
	url = {https://www.light-am.com/article/doi/10.37188/lam.2022.043},
	doi = {10.37188/lam.2022.043},
	language = {en},
	number = {4},
	urldate = {2025-07-25},
	journal = {Light: Advanced Manufacturing},
	author = {Petrov, Nikolay V. and Sokolenko, Bogdan and Kulya, Maksim S. and Gorodetsky, Andrei and Chernykh, Aleksey V.},
	year = {2022},
	note = {Publisher: Changchun Institute of Optics, Fine Mechanics and Physics, Chinese Academy of Sciences},
	pages = {1},
}

@article{gong_research_2023,
	title = {Research progress on terahertz achromatic broadband polarization wave plates},
	volume = {166},
	copyright = {https://www.elsevier.com/tdm/userlicense/1.0/},
	issn = {0030-3992},
	url = {https://linkinghub.elsevier.com/retrieve/pii/S0030399223005261},
	doi = {10.1016/j.optlastec.2023.109633},
	language = {en},
	urldate = {2025-07-25},
	journal = {Optics \& Laser Technology},
	author = {Gong, Yandong and Zhang, Zhuo and Tang, Jianxiong and Ma, Lan and Pang, Kai},
	month = nov,
	year = {2023},
	note = {Publisher: Elsevier BV},
	pages = {109633},
}

@article{zhang_review_2024,
	title = {A review of terahertz wave plate on metasurface},
	copyright = {https://www.springernature.com/gp/researchers/text-and-data-mining},
	issn = {0972-8821, 0974-6900},
	url = {https://link.springer.com/10.1007/s12596-024-02268-0},
	doi = {10.1007/s12596-024-02268-0},
	language = {en},
	urldate = {2025-07-25},
	journal = {Journal of Optics},
	author = {Zhang, Ji and Gong, Yandong},
	month = sep,
	year = {2024},
	note = {Publisher: Springer Science and Business Media LLC},
}

@article{dong_polarization_2009,
	title = {Polarization state and {Mueller} matrix measurements in terahertz-time domain spectroscopy},
	volume = {282},
	copyright = {https://www.elsevier.com/tdm/userlicense/1.0/},
	issn = {0030-4018},
	url = {https://linkinghub.elsevier.com/retrieve/pii/S0030401809005938},
	doi = {10.1016/j.optcom.2009.06.035},
	language = {en},
	number = {18},
	urldate = {2025-07-25},
	journal = {Optics Communications},
	author = {Dong, Hui and Gong, Yandong and Paulose, Varghese and Hong, Minghui},
	month = sep,
	year = {2009},
	note = {Publisher: Elsevier BV},
	pages = {3671--3675},
}

@article{dong_measurement_2010,
	title = {Measurement of {Stokes} parameters of terahertz radiation in terahertz time‐domain spectroscopy},
	volume = {52},
	copyright = {http://onlinelibrary.wiley.com/termsAndConditions\#vor},
	issn = {0895-2477, 1098-2760},
	url = {https://onlinelibrary.wiley.com/doi/10.1002/mop.25450},
	doi = {10.1002/mop.25450},
	language = {en},
	number = {10},
	urldate = {2025-07-25},
	journal = {Microwave and Optical Technology Letters},
	author = {Dong, Hui and Gong, Yandong and Olivo, Malini},
	month = oct,
	year = {2010},
	note = {Publisher: Wiley},
	pages = {2319--2324},
}

@article{gong_cross-polarization_2011,
	title = {Cross-polarization {Response} of a {Two}-contact {Photoconductive} {Terahertz} {Detector}},
	language = {en},
	author = {Gong, Yandong and Dong, Hui and Chen, Zhining},
	year = {2011},
}

@article{neshat_improved_2012,
	title = {Improved measurement of polarization state in terahertz polarization spectroscopy},
	volume = {37},
	copyright = {https://doi.org/10.1364/OA\_License\_v1\#VOR},
	issn = {0146-9592, 1539-4794},
	url = {https://opg.optica.org/abstract.cfm?URI=ol-37-11-1811},
	doi = {10.1364/ol.37.001811},
	language = {en},
	number = {11},
	urldate = {2025-07-25},
	journal = {Optics Letters},
	author = {Neshat, M. and Armitage, N. P.},
	month = jun,
	year = {2012},
	note = {Publisher: Optica Publishing Group},
	pages = {1811},
}

@article{pfleger_advanced_2014,
	title = {Advanced birefringence measurements in standard terahertz time-domain spectroscopy},
	volume = {53},
	copyright = {https://doi.org/10.1364/OA\_License\_v1\#VOR},
	issn = {1559-128X, 2155-3165},
	url = {https://opg.optica.org/abstract.cfm?URI=ao-53-15-3183},
	doi = {10.1364/ao.53.003183},
	language = {en},
	number = {15},
	urldate = {2025-07-25},
	journal = {Applied Optics},
	author = {Pfleger, Michael and Roitner, Heinz and Pühringer, Harald and Wiesauer, Karin and Grün, Hubert and Katletz, Stefan},
	month = may,
	year = {2014},
	note = {Publisher: Optica Publishing Group},
	pages = {3183},
}

@article{kan_enantiomeric_2015,
	title = {Enantiomeric switching of chiral metamaterial for terahertz polarization modulation employing vertically deformable {MEMS} spirals},
	volume = {6},
	copyright = {https://creativecommons.org/licenses/by/4.0},
	issn = {2041-1723},
	url = {https://www.nature.com/articles/ncomms9422},
	doi = {10.1038/ncomms9422},
	language = {en},
	number = {1},
	urldate = {2025-07-25},
	journal = {Nature Communications},
	author = {Kan, Tetsuo and Isozaki, Akihiro and Kanda, Natsuki and Nemoto, Natsuki and Konishi, Kuniaki and Takahashi, Hidetoshi and Kuwata-Gonokami, Makoto and Matsumoto, Kiyoshi and Shimoyama, Isao},
	month = oct,
	year = {2015},
	note = {Publisher: Springer Science and Business Media LLC},
}

@article{choi_terahertz_2022,
	title = {Terahertz {Circular} {Dichroism} {Spectroscopy} of {Molecular} {Assemblies} and {Nanostructures}},
	volume = {144},
	copyright = {https://doi.org/10.15223/policy-029},
	issn = {0002-7863, 1520-5126},
	url = {https://pubs.acs.org/doi/10.1021/jacs.2c04817},
	doi = {10.1021/jacs.2c04817},
	language = {en},
	number = {50},
	urldate = {2025-07-25},
	journal = {Journal of the American Chemical Society},
	author = {Choi, Won Jin and Lee, Sang Hyun and Park, Bum Chul and Kotov, Nicholas A.},
	month = dec,
	year = {2022},
	note = {Publisher: American Chemical Society (ACS)},
	pages = {22789--22804},
}

@article{choi_chiral_2022,
	title = {Chiral phonons in microcrystals and nanofibrils of biomolecules},
	volume = {16},
	copyright = {https://www.springernature.com/gp/researchers/text-and-data-mining},
	issn = {1749-4885, 1749-4893},
	url = {https://www.nature.com/articles/s41566-022-00969-1},
	doi = {10.1038/s41566-022-00969-1},
	language = {en},
	number = {5},
	urldate = {2025-07-25},
	journal = {Nature Photonics},
	author = {Choi, Won Jin and Yano, Keiichi and Cha, Minjeong and Colombari, Felippe M. and Kim, Ji-Young and Wang, Yichun and Lee, Sang Hyun and Sun, Kai and Kruger, John M. and De Moura, André F. and Kotov, Nicholas A.},
	month = may,
	year = {2022},
	note = {Publisher: Springer Science and Business Media LLC},
	pages = {366--373},
}

@article{chen_introduction_2022,
	title = {An introduction to terahertz time-domain spectroscopic ellipsometry},
	volume = {7},
	copyright = {https://creativecommons.org/licenses/by/4.0/},
	issn = {2378-0967},
	url = {https://pubs.aip.org/app/article/7/7/071101/2835194/An-introduction-to-terahertz-time-domain},
	doi = {10.1063/5.0094056},
	language = {en},
	number = {7},
	urldate = {2025-07-25},
	journal = {APL Photonics},
	author = {Chen, X. and Pickwell-MacPherson, E.},
	month = jul,
	year = {2022},
	note = {Publisher: AIP Publishing},
}

@article{yuan_terahertz_2021,
	title = {Terahertz dual-band polarization control and wavefront shaping over freestanding dielectric binary gratings with high efficiency},
	volume = {143},
	copyright = {https://www.elsevier.com/tdm/userlicense/1.0/},
	issn = {0143-8166},
	url = {https://linkinghub.elsevier.com/retrieve/pii/S0143816621001068},
	doi = {10.1016/j.optlaseng.2021.106636},
	language = {en},
	urldate = {2025-07-25},
	journal = {Optics and Lasers in Engineering},
	author = {Yuan, Yiwu and Cheng, Jierong and Dong, Xipu and Fan, Fei and Wang, Xianghui and Chang, Shengjiang},
	month = aug,
	year = {2021},
	note = {Publisher: Elsevier BV},
	pages = {106636},
}

@article{jackel_achromatic_2022,
	title = {Achromatic {Quarter}-{Waveplate} for the {Terahertz} {Frequency} {Range} {Made} by {3D} {Printing}},
	volume = {43},
	copyright = {https://creativecommons.org/licenses/by/4.0},
	issn = {1866-6892, 1866-6906},
	url = {https://link.springer.com/10.1007/s10762-022-00870-6},
	doi = {10.1007/s10762-022-00870-6},
	language = {en},
	number = {7-8},
	urldate = {2025-07-25},
	journal = {Journal of Infrared, Millimeter, and Terahertz Waves},
	author = {Jäckel, Alexander and Ulm, David and Kleine-Ostmann, Thomas and Castro-Camus, Enrique and Koch, Martin and Ornik, Jan},
	month = aug,
	year = {2022},
	note = {Publisher: Springer Science and Business Media LLC},
	pages = {573--581},
}

@article{guan_terahertz_2022,
	title = {Terahertz wideband quarter wave plate based on freestanding grating structures},
	volume = {61},
	issn = {0091-3286},
	url = {https://www.spiedigitallibrary.org/journals/optical-engineering/volume-61/issue-12/125101/Terahertz-wideband-quarter-wave-plate-based-on-freestanding-grating-structures/10.1117/1.OE.61.12.125101.full},
	doi = {10.1117/1.oe.61.12.125101},
	language = {en},
	number = {12},
	urldate = {2025-07-25},
	journal = {Optical Engineering},
	author = {Guan, Shengnan and Cheng, Jierong and Yuan, Yiwu and Dong, Xipu and Fan, Fei and Wang, Xianghui and Chang, Shengjiang},
	month = dec,
	year = {2022},
	note = {Publisher: SPIE-Intl Soc Optical Eng},
}

@article{wu_ultrabroadband_2022,
	title = {Ultra‐{Broadband} {Terahertz} {Polarization} {Conversion} {Enabled} by {All}‐{Dielectric} {Grating} {Structures}},
	volume = {3},
	copyright = {http://creativecommons.org/licenses/by/4.0/},
	issn = {2699-9293, 2699-9293},
	url = {https://onlinelibrary.wiley.com/doi/10.1002/adpr.202200033},
	doi = {10.1002/adpr.202200033},
	language = {en},
	number = {10},
	urldate = {2025-07-25},
	journal = {Advanced Photonics Research},
	author = {Wu, Liang and Zhang, Xi and Fu, Yi and Kang, Kai and Ding, Xin and Yao, Jianquan and Wang, Zhiyong and Han, Jiaguang and Zhang, Weili},
	month = oct,
	year = {2022},
	note = {Publisher: Wiley},
}

@article{jiang_ultrawide_2023,
	title = {Ultrawide tunable terahertz phase shifter based on a double-layer liquid crystal–dielectric grating},
	volume = {40},
	copyright = {https://doi.org/10.1364/OA\_License\_v2\#VOR},
	issn = {0740-3224, 1520-8540},
	url = {https://opg.optica.org/abstract.cfm?URI=josab-40-10-2650},
	doi = {10.1364/josab.497577},
	language = {en},
	number = {10},
	urldate = {2025-07-25},
	journal = {Journal of the Optical Society of America B},
	author = {Jiang, Songlin and Fan, Fei and Ji, Yunyun and Zhao, Huijun and Cheng, Jierong and Wang, Xianghui and Chang, Shengjiang},
	month = oct,
	year = {2023},
	note = {Publisher: Optica Publishing Group},
	pages = {2650},
}

@article{zhang_analysis_2024,
	title = {Analysis of a {THz} unit-structured grating metasurface wave plate},
	volume = {63},
	copyright = {https://doi.org/10.1364/OA\_License\_v2\#VOR},
	issn = {1559-128X, 2155-3165},
	url = {https://opg.optica.org/abstract.cfm?URI=ao-63-25-6581},
	doi = {10.1364/ao.537382},
	language = {en},
	number = {25},
	urldate = {2025-07-25},
	journal = {Applied Optics},
	author = {Zhang, Ji and Gong, Yandong},
	month = sep,
	year = {2024},
	note = {Publisher: Optica Publishing Group},
	pages = {6581},
}

@article{xu_dispersioncompensated_2024,
	title = {Dispersion‐{Compensated} {Terahertz} {Ultra}‐{Broadband} {Quarter} and {Half} {Wave} {Plates} in a {Dielectric}‐{Metal} {Hybrid} {Metadevice}},
	volume = {12},
	copyright = {http://onlinelibrary.wiley.com/termsAndConditions\#vor},
	issn = {2195-1071, 2195-1071},
	url = {https://onlinelibrary.wiley.com/doi/10.1002/adom.202302696},
	doi = {10.1002/adom.202302696},
	language = {en},
	number = {13},
	urldate = {2025-07-25},
	journal = {Advanced Optical Materials},
	author = {Xu, Shi‐Tong and Zhang, Hui‐Fang and Cong, Longqing and Xue, Zhanqiang and Lu, Dan and Wang, Ying‐Hua and Hu, Xiaofei and Liang, Lanju and Ji, Yunyun and Fan, Fei and Chang, Sheng‐Jiang},
	month = may,
	year = {2024},
	note = {Publisher: Wiley},
}

@article{ayyagari_broadband_2024,
	title = {Broadband high-contrast-grating-type waveplates for the terahertz range},
	volume = {32},
	copyright = {https://creativecommons.org/licenses/by/4.0/},
	issn = {1094-4087},
	url = {https://opg.optica.org/abstract.cfm?URI=oe-32-9-15870},
	doi = {10.1364/oe.521532},
	language = {en},
	number = {9},
	urldate = {2025-07-25},
	journal = {Optics Express},
	author = {Ayyagari, Surya Revanth and Klein, Andreas K. and Indrišiūnas, Simonas and Janonis, Vytautas and Pashnev, Daniil and Mohammed, Abdu Subahan and Ducournau, Guillaume and Stöhr, Andreas and Kašalynas, Irmantas},
	month = apr,
	year = {2024},
	note = {Publisher: Optica Publishing Group},
	pages = {15870},
}

@article{zhang_design_2025,
	title = {Design and analysis of terahertz trilayer achromatic grating metasurface wave plate},
	volume = {15},
	copyright = {https://creativecommons.org/licenses/by-nc-nd/4.0},
	issn = {2045-2322},
	url = {https://www.nature.com/articles/s41598-025-90193-y},
	doi = {10.1038/s41598-025-90193-y},
	language = {en},
	number = {1},
	urldate = {2025-07-25},
	journal = {Scientific Reports},
	author = {Zhang, Ji and Ma, Lan and Gong, Yandong},
	month = feb,
	year = {2025},
	note = {Publisher: Springer Science and Business Media LLC},
}

@article{yuan_high-transmission_2025,
	title = {High-transmission linear-to-circular polarization converter based on a polymer relief grating structure},
	volume = {33},
	copyright = {https://doi.org/10.1364/OA\_License\_v2\#VOR-OA},
	issn = {1094-4087},
	url = {https://opg.optica.org/abstract.cfm?URI=oe-33-8-18573},
	doi = {10.1364/oe.555792},
	language = {en},
	number = {8},
	urldate = {2025-07-25},
	journal = {Optics Express},
	author = {Yuan, Yuan and Guan, Lei and Zeng, Qingling and Kong, Depeng and Wang, Lili},
	month = apr,
	year = {2025},
	note = {Publisher: Optica Publishing Group},
	pages = {18573},
}

@article{saha_imprinted_2010,
	title = {Imprinted quarter wave plate at terahertz frequency},
	volume = {28},
	issn = {2166-2746, 2166-2754},
	url = {https://pubs.aip.org/jvb/article/28/6/C6M83/582399/Imprinted-quarter-wave-plate-at-terahertz},
	doi = {10.1116/1.3497023},
	language = {en},
	number = {6},
	urldate = {2025-07-25},
	journal = {Journal of Vacuum Science \& Technology B, Nanotechnology and Microelectronics: Materials, Processing, Measurement, and Phenomena},
	author = {Saha, Shimul C. and Ma, Yong and Grant, James P. and Khalid, A. and Cumming, David R. S.},
	month = nov,
	year = {2010},
	note = {Publisher: American Vacuum Society},
	pages = {C6M83--C6M87},
}

@article{sun_achromatic_2012,
	title = {Achromatic terahertz quarter-wave retarder in reflection mode},
	volume = {106},
	copyright = {http://www.springer.com/tdm},
	issn = {0946-2171, 1432-0649},
	url = {http://link.springer.com/10.1007/s00340-011-4723-9},
	doi = {10.1007/s00340-011-4723-9},
	language = {en},
	number = {2},
	urldate = {2025-07-25},
	journal = {Applied Physics B},
	author = {Sun, L. and Lü, Z. and Zhang, D. and Zhao, Z. and Yuan, J.},
	month = feb,
	year = {2012},
	note = {Publisher: Springer Science and Business Media LLC},
	pages = {393--398},
}

@article{zhang_thin-form_2013,
	title = {Thin-form birefringence quarter-wave plate for lower terahertz range based on silicon grating},
	volume = {52},
	issn = {0091-3286},
	url = {http://opticalengineering.spiedigitallibrary.org/article.aspx?doi=10.1117/1.OE.52.3.030502},
	doi = {10.1117/1.oe.52.3.030502},
	language = {en},
	number = {3},
	urldate = {2025-07-25},
	journal = {Optical Engineering},
	author = {Zhang, Banghong and Gong, Yandong and Dong, Hui},
	month = feb,
	year = {2013},
	note = {Publisher: SPIE-Intl Soc Optical Eng},
	pages = {030502},
}

@article{zhang_achromatic_2015-1,
	title = {Achromatic terahertz quarter waveplate based on silicon grating},
	volume = {23},
	copyright = {https://doi.org/10.1364/OA\_License\_v1\#VOR-OA},
	issn = {1094-4087},
	url = {https://opg.optica.org/abstract.cfm?URI=oe-23-11-14897},
	doi = {10.1364/oe.23.014897},
	language = {en},
	number = {11},
	urldate = {2025-07-25},
	journal = {Optics Express},
	author = {Zhang, Banghong and Gong, Yandong},
	month = jun,
	year = {2015},
	note = {Publisher: Optica Publishing Group},
	pages = {14897},
}

@article{chen_artificial_2016,
	title = {Artificial high birefringence in all-dielectric gradient grating for broadband terahertz waves},
	volume = {6},
	copyright = {https://creativecommons.org/licenses/by/4.0},
	issn = {2045-2322},
	url = {https://www.nature.com/articles/srep38562},
	doi = {10.1038/srep38562},
	language = {en},
	number = {1},
	urldate = {2025-07-25},
	journal = {Scientific Reports},
	author = {Chen, Meng and Fan, Fei and Xu, Shi-Tong and Chang, Sheng-Jiang},
	month = dec,
	year = {2016},
	note = {Publisher: Springer Science and Business Media LLC},
}

@article{mu_broadband_2019,
	title = {Broadband phase shift engineering for terahertz waves based on dielectric metasurface},
	volume = {434},
	copyright = {https://www.elsevier.com/tdm/userlicense/1.0/},
	issn = {0030-4018},
	url = {https://linkinghub.elsevier.com/retrieve/pii/S0030401818309143},
	doi = {10.1016/j.optcom.2018.10.039},
	language = {en},
	urldate = {2025-07-25},
	journal = {Optics Communications},
	author = {Mu, Qianyi and Lin, Hengzhi and Fan, Fei and Cheng, Jierong and Wang, Xianghui and Chang, Shengjiang},
	month = mar,
	year = {2019},
	note = {Publisher: Elsevier BV},
	pages = {12--18},
}

@article{dong_wideband_2019,
	title = {Wideband sub-{THz} half-wave plate using {3D}-printed low-index metagratings with superwavelength lattice},
	volume = {27},
	copyright = {https://doi.org/10.1364/OA\_License\_v1\#VOR-OA},
	issn = {1094-4087},
	url = {https://opg.optica.org/abstract.cfm?URI=oe-27-1-202},
	doi = {10.1364/oe.27.000202},
	language = {en},
	number = {1},
	urldate = {2025-07-25},
	journal = {Optics Express},
	author = {Dong, Xi-Pu and Cheng, Jie-Rong and Fan, Fei and Xu, Shi-Tong and Wang, Xiang-Hui and Chang, Sheng-Jiang},
	month = jan,
	year = {2019},
	note = {Publisher: Optica Publishing Group},
	pages = {202},
}

@article{rohrbach_3d-printed_2021,
	title = {{3D}-printed {THz} wave- and phaseplates},
	volume = {29},
	copyright = {https://doi.org/10.1364/OA\_License\_v1\#VOR-OA},
	issn = {1094-4087},
	url = {https://opg.optica.org/abstract.cfm?URI=oe-29-17-27160},
	doi = {10.1364/oe.433881},
	language = {en},
	number = {17},
	urldate = {2025-07-25},
	journal = {Optics Express},
	author = {Rohrbach, David and Kang, Bong Joo and Feurer, Thomas},
	month = aug,
	year = {2021},
	note = {Publisher: Optica Publishing Group},
	pages = {27160},
}

@article{delacroix_design_2012,
	title = {Design, manufacturing, and performance analysis of mid-infrared achromatic half-wave plates with diamond subwavelength gratings},
	volume = {51},
	copyright = {https://doi.org/10.1364/OA\_License\_v1\#VOR},
	issn = {1559-128X, 2155-3165},
	url = {https://opg.optica.org/abstract.cfm?URI=ao-51-24-5897},
	doi = {10.1364/ao.51.005897},
	language = {en},
	number = {24},
	urldate = {2025-07-25},
	journal = {Applied Optics},
	author = {Delacroix, Christian and Forsberg, Pontus and Karlsson, Mikael and Mawet, Dimitri and Absil, Olivier and Hanot, Charles and Surdej, Jean and Habraken, Serge},
	month = aug,
	year = {2012},
	note = {Publisher: Optica Publishing Group},
	pages = {5897},
}

@article{mutlu_experimental_2012,
	title = {Experimental realization of a high-contrast grating based broadband quarter-wave plate},
	volume = {20},
	copyright = {https://doi.org/10.1364/OA\_License\_v1\#VOR-OA},
	issn = {1094-4087},
	url = {https://opg.optica.org/oe/abstract.cfm?uri=oe-20-25-27966},
	doi = {10.1364/oe.20.027966},
	language = {en},
	number = {25},
	urldate = {2025-07-25},
	journal = {Optics Express},
	author = {Mutlu, Mehmet and Akosman, Ahmet E. and Kurt, Gokhan and Gokkavas, Mutlu and Ozbay, Ekmel},
	month = dec,
	year = {2012},
	note = {Publisher: Optica Publishing Group},
	pages = {27966},
}

@article{liu_highly_2019,
	title = {Highly {Efficient} {Broadband} {Wave} {Plates} {Using} {Dispersion}-{Engineered} {High}-{Index}-{Contrast} {Subwavelength} {Gratings}},
	volume = {11},
	copyright = {https://link.aps.org/licenses/aps-default-license},
	issn = {2331-7019},
	url = {https://link.aps.org/doi/10.1103/PhysRevApplied.11.064005},
	doi = {10.1103/physrevapplied.11.064005},
	language = {en},
	number = {6},
	urldate = {2025-07-25},
	journal = {Physical Review Applied},
	author = {Liu, Wenxing and Yu, Tianbao and Sun, Yong and Lai, Zhenquan and Liao, Qinghua and Wang, Tongbiao and Yu, Longkun and Chen, Hong},
	month = jun,
	year = {2019},
	note = {Publisher: American Physical Society (APS)},
}

@article{nordin_broadband_1999,
	title = {Broadband form birefringent quarter-wave plate for the mid-infrared wavelength region},
	volume = {5},
	copyright = {https://doi.org/10.1364/OA\_License\_v1\#VOR-OA},
	issn = {1094-4087},
	url = {https://opg.optica.org/oe/abstract.cfm?uri=oe-5-8-163},
	doi = {10.1364/oe.5.000163},
	language = {en},
	number = {8},
	urldate = {2025-07-25},
	journal = {Optics Express},
	author = {Nordin, Gregory and Deguzman, Panfilo},
	month = oct,
	year = {1999},
	note = {Publisher: Optica Publishing Group},
	pages = {163},
}

@article{yi_broadband_2003,
	title = {Broadband achromatic phase retarder by subwavelength grating},
	volume = {227},
	copyright = {https://www.elsevier.com/tdm/userlicense/1.0/},
	issn = {0030-4018},
	url = {https://linkinghub.elsevier.com/retrieve/pii/S0030401803019527},
	doi = {10.1016/j.optcom.2003.09.026},
	language = {en},
	number = {1-3},
	urldate = {2025-07-25},
	journal = {Optics Communications},
	author = {Yi, De-Er and Yan, Ying-Bai and Liu, Hai-Tao and {Si-Lu} and Jin, Guo-Fan},
	month = nov,
	year = {2003},
	note = {Publisher: Elsevier BV},
	pages = {49--55},
}

@article{park_broadband_2024,
	title = {Broadband metamaterial polarizers with high extinction ratio for high-precision terahertz spectroscopic polarimetry},
	volume = {9},
	copyright = {https://creativecommons.org/licenses/by-nc-nd/4.0/},
	issn = {2378-0967},
	url = {https://pubs.aip.org/app/article/9/11/110807/3319393/Broadband-metamaterial-polarizers-with-high},
	doi = {10.1063/5.0228119},
	language = {en},
	number = {11},
	urldate = {2025-07-25},
	journal = {APL Photonics},
	author = {Park, H. and Park, H. and Lee, J. and Shim, J. and Son, H. and Park, J. and Baek, S. and Kim, T.-T.},
	month = nov,
	year = {2024},
	note = {Publisher: AIP Publishing},
}

@inproceedings{qiu_terahertz_2018,
	address = {Beijing, China},
	title = {Terahertz microstructure for artificial birefringence and its applications},
	url = {https://www.spiedigitallibrary.org/conference-proceedings-of-spie/10826/2324953/Terahertz-microstructure-for-artificial-birefringence-and-its-applications/10.1117/12.2324953.full},
	doi = {10.1117/12.2324953},
	language = {en},
	urldate = {2025-07-25},
	booktitle = {Infrared, {Millimeter}-{Wave}, and {Terahertz} {Technologies} {V}},
	publisher = {SPIE},
	author = {Qiu, Jiang and Mu, Qian-Yi and Chang, Shengjiang and Fan, Fei},
	editor = {Zhang, Xi-Cheng and Tani, Masahiko and Zhang, Cunlin},
	month = nov,
	year = {2018},
	pages = {14},
}

@article{zhu_extend_2024,
	title = {Extend the {Bandwidth} of {Terahertz} {Polarization} {Convertors} {Further} via {All}-{Dielectric} {Grating} {Structures}},
	volume = {42},
	copyright = {https://ieeexplore.ieee.org/Xplorehelp/downloads/license-information/IEEE.html},
	issn = {0733-8724, 1558-2213},
	url = {https://ieeexplore.ieee.org/document/10607850/},
	doi = {10.1109/jlt.2024.3432747},
	language = {en},
	number = {23},
	urldate = {2025-07-25},
	journal = {Journal of Lightwave Technology},
	author = {Zhu, Rui and Zou, Die and Xing, Xiaohua and Ding, Xin and Zhang, Guizhong and Yao, Jianquan and Wu, Liang},
	month = dec,
	year = {2024},
	note = {Publisher: Institute of Electrical and Electronics Engineers (IEEE)},
	pages = {8336--8342},
}

@article{xu2020terahertz,
  title={Terahertz time-domain polarimetry (THz-TDP) based on the spinning EO sampling technique: determination of precision and calibration},
  author={Xu, Kuangyi and Bayati, Elyas and Oguchi, Kenichi and Watanabe, Shinichi and Winebrenner, Dale P and Hassan Arbab, M},
  journal={Optics Express},
  volume={28},
  number={9},
  pages={13482--13496},
  year={2020},
  publisher={Optical Society of America}
}

@article{brundrett_homogeneous_1994,
	title = {Homogeneous layer models for high-spatial-frequency dielectric surface-relief gratings: conical diffraction and antireflection designs},
	volume = {33},
	copyright = {https://doi.org/10.1364/OA\_License\_v1\#VOR},
	issn = {0003-6935, 1539-4522},
	shorttitle = {Homogeneous layer models for high-spatial-frequency dielectric surface-relief gratings},
	url = {https://opg.optica.org/abstract.cfm?URI=ao-33-13-2695},
	doi = {10.1364/AO.33.002695},
	language = {en},
	number = {13},
	urldate = {2025-07-27},
	journal = {Applied Optics},
	author = {Brundrett, David L. and Glytsis, Elias N. and Gaylord, Thomas K.},
	month = may,
	year = {1994},
	pages = {2695},
}

@article{brundrett_subwavelength_1996,
	title = {Subwavelength transmission grating retarders for use at 106},
	volume = {35},
	copyright = {https://doi.org/10.1364/OA\_License\_v1\#VOR},
	issn = {0003-6935, 1539-4522},
	url = {https://opg.optica.org/abstract.cfm?URI=ao-35-31-6195},
	doi = {10.1364/AO.35.006195},
	language = {en},
	number = {31},
	urldate = {2025-07-27},
	journal = {Applied Optics},
	author = {Brundrett, D. L. and Glytsis, E. N. and Gaylord, T. K.},
	month = nov,
	year = {1996},
	pages = {6195},
}

@article{rytov_electromagnetic_1956,
	title = {Electromagnetic properties of a finely stratified medium},
	volume = {2},
	journal = {Soviet Physics JEPT},
	author = {Rytov, S},
	year = {1956},
	pages = {466--475},
}

@article{kikuta_ability_1995,
	title = {Ability and limitation of effective medium theory for subwavelength gratings},
	volume = {2},
	number = {2},
	journal = {Optical Review},
	author = {Kikuta, Hisao and Yoshida, Hideo and Iwata, Koichi},
	year = {1995},
	note = {Publisher: Japan Society of Applied Physics},
	pages = {92--99},
}

@article{lalanne_artificial_2003,
	title = {Artificial media optical properties-subwavelength scale},
	author = {Lalanne, Philippe and Hutley, Mike},
	year = {2003},
	note = {Pages: 62–71
Publication Title: Encyclopedia of optical engineering
Volume: 1},
}

@book{noauthor_rf_nodate,
	title = {{RF} {Module} {User}’s {Guide}, version 6.3, {COMSOL}, {Inc}.},
	url = {https://doc.comsol.com/6.3/doc/com.comsol.help.rf/RFModuleUsersGuide.pdf},
	urldate = {2025-07-29},
}

@article{byrnes2016multilayer,
  title={Multilayer optical calculations},
  author={Byrnes, Steven J},
  journal={arXiv preprint arXiv:1603.02720},
  year={2016}
}

@article{jepsen2019phase,
  title={Phase retrieval in terahertz time-domain measurements: a “how to” tutorial},
  author={Jepsen, Peter Uhd},
  journal={Journal of Infrared, Millimeter, and Terahertz Waves},
  volume={40},
  number={4},
  pages={395--411},
  year={2019},
  publisher={Springer}
}

\end{document}